\documentclass[preprint2,twocolumn]{aastex631}

\usepackage{amsmath,graphicx}
\usepackage{bm}
\usepackage{multirow}
\usepackage{booktabs}
\usepackage{fix-cm}
\usepackage{tablefootnote}
\usepackage{relsize}
\usepackage{tikz}
\usetikzlibrary{spy}
\shorttitle{Robust R2D2}
\shortauthors{Aghabiglou et al.}
\graphicspath{{./}{figures/}}
\newcommand{\xb}{\ensuremath{\boldsymbol{x}}}

\newcommand{\rb}{\ensuremath{\boldsymbol{r}}}
\newcommand{\Nb}{\ensuremath{\boldsymbol{\mathsf{N}}}}
\definecolor{RED}{named}{red}
\newcommand{\thetab}{\ensuremath{\boldsymbol{\theta}}}
\newcommand{\bmu}{\ensuremath{\boldsymbol{\mu}}}
\newcommand{\bsigma}{\ensuremath{\boldsymbol{\sigma}}}

\begin{document}
\title{Toward a Robust R2D2 Paradigm for Radio-interferometric Imaging: Revisiting Deep Neural Network Training and Architecture}


\correspondingauthor{Yves Wiaux}
\email{y.wiaux@hw.ac.uk}

\author[0000-0001-6024-649X]{Amir Aghabiglou}
\affiliation{Institute of Sensors, Signals and Systems, Heriot-Watt University, Edinburgh EH14 4AS, United Kingdom}

\author[0009-0004-1056-5619]{Chung San Chu}
\affiliation{Institute of Sensors, Signals and Systems, Heriot-Watt University, Edinburgh EH14 4AS, United Kingdom}

\author[0009-0001-3611-2229]{Chao Tang}
\affiliation{Institute of Sensors, Signals and Systems, Heriot-Watt University, Edinburgh EH14 4AS, United Kingdom}
\affiliation{EPCC, University of Edinburgh, Potterrow, Edinburgh EH8 9BT, United Kingdom}

\author[0000-0002-7903-3619]{Arwa Dabbech}
\affiliation{Institute of Sensors, Signals and Systems, Heriot-Watt University, Edinburgh EH14 4AS, United Kingdom}

\author[0000-0002-1658-0121]{Yves Wiaux}
\affiliation{Institute of Sensors, Signals and Systems, Heriot-Watt University, Edinburgh EH14 4AS, United Kingdom}

\begin{abstract}
The R2D2 Deep Neural Network (DNN) series was recently introduced for image formation in radio interferometry. 
It can be understood as a learned version of CLEAN, whose minor cycles are substituted with DNNs. We revisit R2D2 on the grounds of series convergence, training methodology, and DNN architecture, improving its robustness in terms of generalizability beyond training conditions, capability to deliver high data fidelity, and epistemic uncertainty. First, while still focusing on telescope-specific training, we enhance the learning process by randomizing Fourier sampling integration times, incorporating multiscan multinoise configurations, 
and varying imaging settings, including pixel resolution and visibility-weighting scheme. 
Second, we introduce a convergence criterion whereby the reconstruction process stops when the data residual is compatible with noise, rather than simply using all available DNNs. This not only increases the reconstruction efficiency by reducing its computational cost, but also refines training by pruning out the data/image pairs for which optimal data fidelity is reached before training the next DNN. 
Third, we substitute R2D2's early U-Net DNN with a novel architecture (U-WDSR) combining U-Net and WDSR, which leverages wide activation, dense skip connections, weight normalization, and low-rank convolution to improve feature reuse and reconstruction precision. 
As previously, R2D2 was trained for monochromatic intensity imaging with the Very Large Array at fixed $512 \times 512$ image size. Simulations on a wide range of inverse problems and a case study on real data reveal that the new R2D2 model consistently outperforms its earlier version in image reconstruction quality, data fidelity, and epistemic uncertainty.
\end{abstract}

\keywords{Computational methods (1965) --- Neural networks (1933) --- Astronomy image processing (2306) --- Aperture synthesis (53)}


\section{Introduction}
Radio Interferometry (RI) is a core data acquisition modality in radio astronomy that enables the study of intricate phenomena in the Universe, such as cosmic magnetic fields, galaxy formation, and the properties of black holes.
The advent of advanced radio telescopes, such as MeerKAT \citep{jonas2016proc}, the Australian Square Kilometre Array Pathfinder \citep[ASKAP;][]{hotan2021}, the Low-Frequency Array \citep[LOFAR;][]{van2013lofar}, and the upcoming Square Kilometre Array \citep[SKA; ][]{Labate2022,Swart22}, has pushed the field forward, offering unprecedented resolution and sensitivity. However, these advancements pose challenges for the image formation process, now due to scale to large data volumes to form densely populated radio images at the target resolution and dynamic range.

RI data consists of noisy, undersampled Fourier measurements of the target radio image. The underpinning image formation problem is an ill-posed inverse problem. Thanks to its simplicity and computational efficiency, the CLEAN algorithm~\citep{hogbom1974aperture} has been a longstanding standard in RI imaging. However, CLEAN’s limitations become apparent when addressing complex emission and high dynamic ranges. The algorithm's reliance on a simplistic prior model can lead to suboptimal results, particularly when the required angular resolution surpasses the nominal instrumental resolution.

In response to these limitations, the field has shifted toward more advanced computational imaging techniques. Algorithms grounded in optimization theory, such as the SARA family~\citep{carrillo2012sparsity,onose2016scalable,onose2017accelerated,repetti2020forward,terris2022}, have demonstrated superior image reconstruction capabilities by incorporating handcrafted sparsity-based regularization, enabling higher resolution and more physical reconstruction of the target signal than CLEAN. Despite their high image precision, these algorithms remain highly iterative at the target high dynamic ranges, which leads to inevitable computational limitations in large-scale regimes.

More recently, the integration of deep learning into image reconstruction has opened new avenues for enhancing both speed and precision. On the one hand, end-to-end DNNs, promising ultra-fast reconstructions, have been explored, albeit with trade-offs in robustness, generalizability, and interpretability \citep{connor2022deep,geyer2023deep}. On the other hand, plug-and-play (PnP) algorithms, such as AIRI \citep{terris2022,terris2025airi}, combine the strengths of deep learning and optimization, offering a flexible framework by replacing regularization terms with learned denoisers. These hybrid algorithms are also highly iterative in nature, raising concerns about their computational efficiency.

Very recently, we have introduced the Residual-to-Residual
DNN series for high-Dynamic-range imaging-paradigm \citep[R2D2; ][]{aghabiglou2023a,aghabiglou2024r2d2,dabbech2024cleaning}, aiming to improve both precision and computational efficiency over the state of the art. R2D2 forms an image as a series of residual images iteratively estimated as outputs of DNNs, taking the previous iteration and associated data residual as inputs. 
The first incarnation of the R2D2 algorithm was underpinned by the U-Net architecture. 
Despite its promising precision and computational efficiency in both simulation and real data, R2D2's robustness across diverse imaging settings was unexplored, including varying visibility-weighting schemes, pixel resolution, and image sizes. Generalizing the approach from the current monochromatic intensity imaging setting to address wideband polarization imaging is yet to be investigated.

In this paper, we build on these foundations and propose several key advancements to address the limitations of R2D2 while maintaining its focus on monochromatic intensity imaging with the Very Large Array (VLA) at an image size of $512 \times 512$ pixels. These include training methodology, convergence criterion, and DNN architecture. Our contributions aim to improve R2D2’s robustness, defined in terms of generalizability beyond training conditions, capability to deliver high data fidelity, and epistemic uncertainty. 

Firstly, we generalize the training setup from \cite{aghabiglou2024r2d2}, by introducing stochastic variations in key observational and imaging parameters. In our previous study, we adopted fixed imaging settings whereby (i) the imaging pixel resolution was set to enable a fixed ratio of the imaging resolution to the nominal instrumental resolution, and (ii) the Briggs weighting scheme was applied with a fixed value of the robustness parameter that controls the trade-off between uniform and natural weighting schemes. Other observation settings such as the integration time were also fixed. In this work, we randomize all the above parameters.
Additionally, we extend the algorithm to support multinoise and multiscan configurations, enabling it to handle more complex and 
realistic observational scenarios. Secondly, we introduce a convergence criterion whereby the reconstruction process is deemed complete and iterations stop when the data residual is compatible with noise, rather than utilizing all available DNNs. This not only reduces the training computational cost but also improves reconstruction efficiency. Concurrently, a dynamic data-pruning procedure 
is applied during training to both the training and validation sets, enhancing overall training computational efficiency and model learning. Thirdly, we propose a novel DNN architecture as core architecture for R2D2, dubbed U-WDSR \citep{aghabiglou2023a}, which combines the strengths of the U-Net architecture with WDSR residual blocks \citep{yu2018wide}. The advanced architecture enables the recovery of finer details with enhanced data fidelity.

Furthermore, we provide a comprehensive evaluation of these contributions by benchmarking R2D2 against state-of-the-art RI imaging algorithms, namely AIRI and uSARA {\citep{terris2022}}. R2D2 is implemented as a fully Python GPU-enabled algorithm. For a fair comparison, we transition both AIRI and uSARA implementations from MATLAB to GPU-enabled Python, significantly improving their computational efficiency. These {GPU-accelerated implementations} are integrated into BASPLib\footnote{BASPLib: The Biomedical and Astronomical Signal Processing library is available at \url{https://basp-group.github.io/BASPLib/}.}, {a publicly available code library dedicated to solving imaging inverse problems.} R2D2 is also benchmarked against multi-scale CLEAN from the widely-used WSClean software \citep{offringa2014,offringa2017}.

The remainder of this paper is organized as follows: Section~\ref{R2D2_Algorithms} revisits the data model for RI imaging and provides an overview of the R2D2 algorithmic structure. Section~\ref{robust_r2d2} delves into the training methodology for robust R2D2 algorithm, detailing the construction of a generalized training set, convergence criterion, the novel U-WDSR DNN architecture, epistemic uncertainty quantification, training implementation and computational cost. Section~\ref{SIMULATIONS&RESULTS} examines R2D2’s robustness and generalizability under diverse experimental setups, and evaluates its performance in comparison with the earlier R2D2 model \citep{aghabiglou2024r2d2} and the benchmark algorithms, with a focus on imaging precision and computational efficiency. Additionally, it explores epistemic uncertainty to further validate R2D2’s robustness. Section~\ref{realdata} revisits real observations of the radio galaxy Cygnus~A with the new R2D2 model. Finally, Section~\ref{Conclusions} summarizes the key findings and provides directions for future work.

\section{R2D2 Paradigm} \label{R2D2_Algorithms}
This section revisits the RI data model in the context of monochromatic intensity imaging and provides an overview of R2D2 algorithmic structure. 

\subsection{RI data model}

Under the assumption of nonpolarized monochromatic radio emission, spanning a narrow field of view, RI data, also called visibilities, are incomplete noisy Fourier measurements of the intensity image of interest. 
Let $\xb^\star \in \mathbb{R}_+^N$ represent the unknown intensity (and thus non-negative) image of the sky, with $N$ pixels. {Formally,} the RI data model reads:
\begin{equation}
\bm{y} = \bm{\Phi x}^\star + \bm{n},
\end{equation}
where $\bm{y} \in \mathbb{C}^M$ is the vector of observed visibilities, and $\bm{n} \in \mathbb{C}^M$ is the additive noise vector, typically assumed to be a complex random Gaussian noise with mean zero and variance $\tau^{2} > 0$. The measurement operator $\bm{\Phi}: \mathbb{R}^N \rightarrow \mathbb{C}^M$ represents the non-uniform Fourier sampling, and {is modelled using the non-uniform fast Fourier transform \citep[NUFFT,][]{Fessler2003}} such that
$\bm{\Phi} = {\textrm{\textbf{G}}} \textrm{\textbf{F}} \textrm{\textbf{Z}}$, where $\textrm{\textbf{G}} \in \mathbb{C}^{M \times D}$ is a sparse interpolation matrix, $\textrm{\textbf{F}} \in \mathbb{C}^{D \times D}$ is the 2D discrete Fourier transform, and $\textrm{\textbf{Z}} \in \mathbb{R}^{D \times N}$ is a zero-padding operator which also includes the correction for the convolution in the Fourier domain through $\textrm{\textbf{G}}$. Often, a visibility-weighting scheme (e.g.~Briggs weighting) is applied to the RI data and injected into the measurement operator model to balance sensitivity and resolution depending on the target science. The current measurement operator model does not account for direction-dependent effects (DDEs), such as the so-called $w$-effect arising from the non-coplanarity of the array, or unknown atmospheric and instrumental perturbations, estimated during a calibration step \citep{smirnov2011}. 
When available, DDE estimates can be easily included as Fourier convolution kernels in the interpolation matrix G \citep{dabbech2021}.

The RI data model can be formulated in the image domain through a normalized back-projection via the adjoint of the measurement operator. More precisely, the back-projected data $\xb_\textrm{d}\in \mathbb{R}^N$, also called the dirty image, is obtained as $\xb_\textrm{d} = \kappa\mathrm{Re}\{\bm{\Phi}^\dagger \bm{y}\}$, 
where $(.^\dagger)$ denotes the adjoint of its argument. The normalization factor $\kappa$ ensures that the peak value of the point spread function (PSF) is equal to one, a conventional normalization in RI imaging. Specifically, $\kappa = {\max \left(\mathrm{Re}\{\bm{\Phi}^\dagger \bm{\Phi} \bm{\delta}\}\right)}^{-1}$, where $\bm{\delta} \in \mathbb{R}^N$ is an image with a value of 1 at its center and 0 elsewhere. 
The operator $\mathrm{Re}\{\cdot\}$ ensures that the image domain representation is real-valued, as expected for intensity images. 

\subsection{Algorithmic structure}

The R2D2 algorithm involves training a collection of $I$ DNNs, denoted as $(\Nb_{\widehat{\thetab}^{(i)}})_{1 \leq i \leq I}$, defined by their learned parameters $({\widehat{\thetab}^{(i)}} \in \mathbb{R}^Q)_{1 \leq i \leq I}$. Each DNN {$\Nb_{\widehat{\thetab}^{(i)}}$} takes as input the previous image estimate $\xb^{(i-1)}$ and its associated residual dirty image $\rb^{(i-1)}$ (i.e.~back-projected data residual) defined as:
\begin{equation}
\label{eq:res_update}
\rb^{(i-1)} = \xb_\textrm{d} - \kappa \mathrm{Re}\{\bm{\Phi}^\dagger \bm{\Phi}\}\xb^{(i-1)}.
\end{equation}
For the first iteration, the initial image estimate is set to zero ($\xb^{(0)} = \bm{0}$) and the associated residual dirty image corresponds to the dirty image ($\rb^{(0)} = \xb_\textrm{d}$). The current image estimate is then updated as:
\begin{equation}
\xb^{(i)} = [\xb^{(i-1)} + \Nb_{\widehat{\thetab}^{(i)}}(\rb^{(i-1)},\xb^{(i-1)})]_{+},\label{eq:update}
\end{equation}
where $[\cdot]_{+}$ denotes the projection of its argument into the non-negative orthant, which corresponds to setting negative pixel values to zero. Ensuring the non-negativity of the reconstructed image is an essential physical constraint on intensity images. Here, each DNN $\Nb_{\widehat{\thetab}^{(i)}}$ learns to predict a residual image using the previous image estimate and its corresponding residual dirty image. The output residual image is then added to the previous image estimate.
The final reconstruction corresponds to the $I$-th iteration {i.e.}~$\widehat{\xb}=\xb^{(I)}$. In the absence of the non-negativity constraint, R2D2's reconstruction would 
take the simple series expression $\widehat{\xb}= \sum_{i=1}^{I} {\Nb_{\widehat{\thetab}^{(i)}}(\rb^{(i-1)},\xb^{(i-1)})}$, which motivates the denomination of the ``DNN series''. Early R2D2 models have shown that the underlying DNNs progressively capture finer details and fainter emission over the iterations. Importantly, by incorporating accurate updates of the back-projected data residuals, R2D2 effectively suppresses hallucination artifacts that are inconsistent with the data. In addition, its DNNs involve an iteration-specific normalization strategy, making them agnostic to varying intensity range and less prone to generalization issues  \citep[see][for details]{aghabiglou2024r2d2}. 

R2D2 DNNs are trained sequentially using supervised learning, whereby at each iteration $i$, the training dataset of the current DNN is updated from the output of the preceding network. More specifically, the training dataset consists of $K$ samples 
 corresponding to the image triplets $(\xb_l^\star,{\xb_k^{(i-1)}},{\rb_k^{(i-1)}})_{1\leq k\leq K}$.
The goal is to minimize the error between the current image estimate $\xb^{(i)}_{k}$ and the target ground-truth image $\xb^{\star}_{k}$ for the $k$-th training sample. This is achieved using an $\ell_1$-norm loss function with a non-negativity constraint on the target image:
\begin{equation}
 \min_{{\thetab}^{(i)}\in \mathbb{R}^Q}\frac{1}{K} \sum_{k=1}^{K} ~ \| {\xb}^{\star}_{k} - [{\xb}^{(i-1)}_{k} + {{\Nb}}_{{\thetab}^{(i)}}(\rb_{k}^{(i-1)}, {\xb}^{(i-1)}_{k})]_{+} \|_{1},\label{eq:r2d2_loss}
\end{equation}
This loss ensures the DNN generates output residual images, promoting the non-negativity of the image estimate, while penalizing large deviations from the ground truth. Loss functions of the form given by Equation \eqref{eq:r2d2_loss} are optimized using the Root Mean Square Propagation (RMSProp) algorithm, with the learnable parameters of each network initialized from the estimated parameters of the preceding network. 

\section{R2D2 robustness} \label{robust_r2d2}
This section describes key features ensuring the robustness of R2D2, targeting the formation of monochromatic intensity images under a VLA-specific observational setup. We focus on three core aspects: the training methodology that improves model generalization, the introduction of a convergence criterion, and a novel DNN architecture underpinning the R2D2 series. For insights into the reliability and interpretability of the algorithm’s outputs, we present an ensemble averaging approach for epistemic uncertainty quantification.

\subsection{Generalized training dataset} \label{trainingset}

In building the training dataset, we followed closely the training setup described in \citet{aghabiglou2024r2d2}. It consists of $K$ pairs of ground-truth images and their corresponding dirty images of size $N=512 \times 512$.
Ground-truth images are derived from low-dynamic-range radio and optical astronomy images as well as medical imaging sources. The latter were included for increased morphological diversity. To avoid introducing bias, dedicated transforms (e.g.~rotation, translation, concatenation, edge smoothing) were applied to deconstruct their anatomical features. All images were denoised, normalized within the range $[0,1]$, and then pixel-wise exponentiated to achieve high-dynamic-range ground-truth images. The dynamic range, denoted by $a$, was randomly selected within the range $[10^3,~ 5\times 10^5]$. Full details can be found in \citet{aghabiglou2024r2d2} alongside examples of raw images and the associated curated ground truth images.
Realistic RI data were simulated, combining VLA configurations A and C. Fourier sampling patterns were generated by uniformly randomizing several parameters including (i) the pointing direction, (ii) the total observation durations with configurations A and C (denoted by $t_{\textrm{obs-A}}$ and $t_{\textrm{obs-C}}$, respectively), and (iii) the spectral specifications. These consist of the frequency bandwidth, described by the ratio of the highest to the lowest frequency $(\rho_{\textrm{freq}})$, and the number of observation frequencies combined for image formation $(n_{\textrm{freq}})$. 

To enhance the robustness of R2D2 to varying observational conditions, the RI Fourier sampling is further diversified in this study by randomizing the previously fixed integration time $(t_{\textrm{samp.}})$ {in the set} $\{4,8,16,32\}$ seconds. The total number of points in the resulting Fourier sampling patterns ranges from $2 \times 10^5$ to $2.7 \times 10^7$, spanning a range approximately one order of magnitude wider than the previously considered patterns. Moreover, a multiscan multinoise setup was considered instead of a single-scan setup. In practice, the target radio source is often observed alongside other nearby calibrator sources with known flux densities. Data acquisition is therefore performed in time scans, alternating between the target source and the calibrator sources for the duration of the observation. The number of time scans $(n_{\textrm{scan}})$ was uniformly randomized between 1 and 8, with a lag time of up to $20\%$ of the observation duration. 
Under these considerations, the standard deviation of the additive noise vector $\bm{n}$ corrupting the simulated RI data $\bm{y}$ varies per time scan and frequency channel. Let $s \in \{1,\dots, n_\textrm{scan}\}$ denote the index of a given time scan, and $f \in \{1,\dots, n_\textrm{freq}\}$ the index of a frequency channel. The standard deviation of the associated noise block $\bm{n}_{s,f}$ denoted by $\tau_{s,f}$ is set following a stipulation of \citet{terris2022} linking the measurement noise to the dynamic range of the radio image of interest. Specifically, $\tau_{s,f}= {a}^{-1} \sqrt{2 \| \mathrm{Re}\{\bm{\Phi}_{s,f}^\dagger \bm{\Phi}_{s,f}\}\|_S} $, where 
$\bm{\Phi}_{s,f}$ is the associated measurement operator block, and $\|. \|_S$ denotes the spectral norm of its argument operator.

\begin{table*}
\centering
\caption{Parameter choice of the training setup $\mathcal{T}_1$ described in \citet{aghabiglou2024r2d2} and the proposed training setup $\mathcal{T}_2$}
\label{table:training_setup}
\hspace{-2.1cm}
\resizebox{1.1\linewidth}{!}
{\begin{tabular}{lccccccccccc}
\hline
\hline
\multirow{3}{*}{Training setup} & \multicolumn{9}{c}{Observational parameters} & \multicolumn{2}{c}{Imaging parameters}               \\ \cline{2-12} 
& DEC   & RA (J2000)   & $t_{\textrm{obs-A}}$                      & $t_{\textrm{obs-C}}$                     & $t_{\textrm{samp.}}$ & \multirow{2}{*}{$\rho_{\textrm{freq}}$}           & \multirow{2}{*}{$n_{\textrm{freq}}$}& \multirow{2}{*}{Noise variance}& \multirow{2}{*}{$n_{\textrm{scan}}$}  & \multirow{2}{*}{$\rho_\textrm{sr}$}                  & \multirow{2}{*}{$\rho_\textrm{br}$}               \\

&(degrees)&(hr)&(hr)&(hr)&(sec.)&&&&\\ 
\hline
$\mathcal{T}_1$                                & \multirow{2}{*}{$[5,60]$} & \multirow{2}{*}{$[0,23]$} & \multirow{2}{*}{$[5,10]$} & \multirow{2}{*}{$[1,3]$} & 36             & \multirow{2}{*}{$[1,2]$} &  \multirow{2}{*}{$\{1,\dots,4\}$}     & homogeneous  &   1            & {1.5}                 & 0             \\  \cline{1-1} \cline{6-6} \cline{9-9} \cline{10-12} 
\rule{0pt}{10pt}
\hspace{-1mm}$\mathcal{T}_2$                                  & {}                                               & {}                                                       & {}                                          &                                          & $\{4, 8, 16, 32\}$              & {}                                 & {}                     & time-scan \& frequency-dependent  & $\{1,\dots,8\}$                  &    $[1.5,2.5]$ & $[-1,1]$ \\ \hline
\end{tabular}
}

\vspace{2mm}
\parbox{\linewidth}{\footnotesize%
\textbf{Note.} Observational parameters include the pointing direction (the declination (DEC) and right ascension (RA)), the total observation time with VLA configurations A ($t_{\textrm{obs-A}}$) and C ($t_{\textrm{obs-C}}$), the sampling integration time $t_{\textrm{samp.}}$, the number of time scans ($n_{\textrm{scan}}$), the frequency bandwidth ratio ($\rho_{\textrm{freq}}$), the number of frequencies ($n_{\textrm{freq}}$), and the properties of the additive random Gaussian noise. Imaging parameters include the super-resolution factor ($\rho_\textrm{sr}$) determining the pixel resolution, and the robustness parameter of Briggs weighting ($\rho_\textrm{br}$). 
Values in $[.,.]$ indicate the lower and upper bounds for generating uniformly random parameter values.
}
\end{table*}

R2D2 robustness to varying imaging settings is also propelled by varying the previously fixed pixel resolution and visibility-weighting scheme adopted for the generation of the dirty images via back-projection. More precisely, Briggs weighting scheme \citep{briggs1995high}, previously adopted with a fixed robustness parameter ($\rho_{\textrm{br}}$), was uniformly randomized in the range $[-1,1]$, with lower values approaching uniform weighting, and higher values approaching natural weighting. The pixel resolution of the dirty images was also randomly chosen to reflect a super-resolution factor during imaging $(\rho_\textrm{sr})$ in the range $[1.5, 2.5]$. In the remainder of this article, we refer to the training setup of \citet{aghabiglou2024r2d2} as $\mathcal{T}_1$, and to the more generalized training setup {proposed herein} as $\mathcal{T}_2$. A summary of the parameter space underlying both $\mathcal{T}_1$ and $\mathcal{T}_2$ is provided in Table~\ref{table:training_setup}.

\subsection{Series convergence} \label{series_convergence}
To improve learning efficiency, in \citet{aghabiglou2024r2d2} we introduced a dynamic pruning strategy applied to the training set. Once an inverse problem is deemed converged, it is removed from the training set for subsequent DNNs in the series. This ensures that each iteration focuses on unsolved problems while reducing computational cost. Convergence of the inverse problem underpinning each training image pair $(\xb^\star_{k},{\xb_{\textrm{d}}}_{k})$ is assessed based on the evolution of the associated residual dirty image. Specifically, it is considered to be solved if, at a given iteration $i > 1$, the residual dirty image satisfies the condition $\| \rb_{k}^{(i)} \|^2_2 \leq \| \kappa_k \mathrm{Re}\{\bm{\Phi}^\dagger_k \bm{\Phi}_k\}\bm{n}_{k} \|^2_2$, where the right-hand side represents the $\ell_2$-norm of the back-projected noise vector in the image-domain, assumed known during training. The sequential training of the series concludes once the evaluation metrics applied to the validation set converge (i.e.~stabilize) or when the size of the training set after pruning falls below an inappropriate size beyond which further training is no longer beneficial. In this study, we extend the pruning strategy to the validation set. 
This extension allows training to continue for more iterations, enabling better reconstruction of unsolved inverse problems, particularly those associated with extreme dynamic-range ground-truth images and low noise levels. Furthermore, we find that applying the entire DNN series to inverse problems pruned during training does not degrade the quality of their reconstructions.

In the image reconstruction step, we also introduce a data-fidelity-based convergence criteria to avoid deploying all terms of the series unnecessarily. In practice, the exact noise level is typically unknown, prompting the use of alternative stopping conditions. The first criterion relies on a user-defined lower bound on the relative variation between consecutive residual dirty images (measured in $\ell_2$-norm), indicating stable data fidelity. In our experiments, the lower bound is set to $10^{-3}$. The second criterion stops the algorithm if the $\ell_2$ norm of the residual dirty image increases twice along the iterations indicating fluctuating data fidelity near convergence. The algorithm stops when either of these criteria is met.

\subsection{U-WDSR DNN architecture}

In this section, we present a novel DNN architecture underpinning the R2D2 algorithm, which combines the WDSR residual body architecture, originally proposed by \citet{yu2018wide} for image and video super-resolution, with the U-Net architecture.  The novel architecture, dubbed U-WDSR, retains the primary structural components of U-Net, including the contracting and expanding paths, skip connections, and pooling/upsampling operations, and incorporates the WDSR residual body as a block interlaced with U-Net's conventional convolution layers. 

\begin{figure*}
    \makebox[\linewidth][l]{\includegraphics[width = 0.7\linewidth]{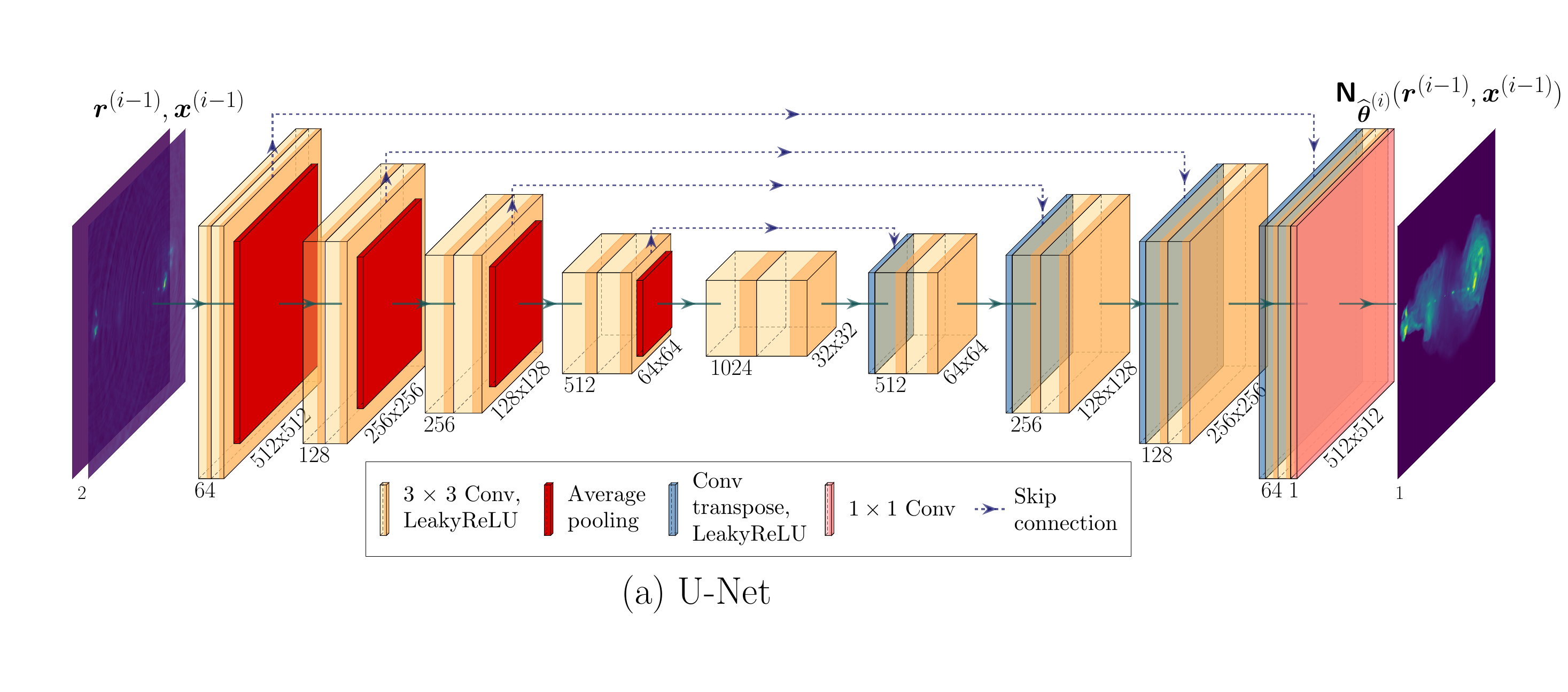}}\\
    \includegraphics[width = \linewidth]{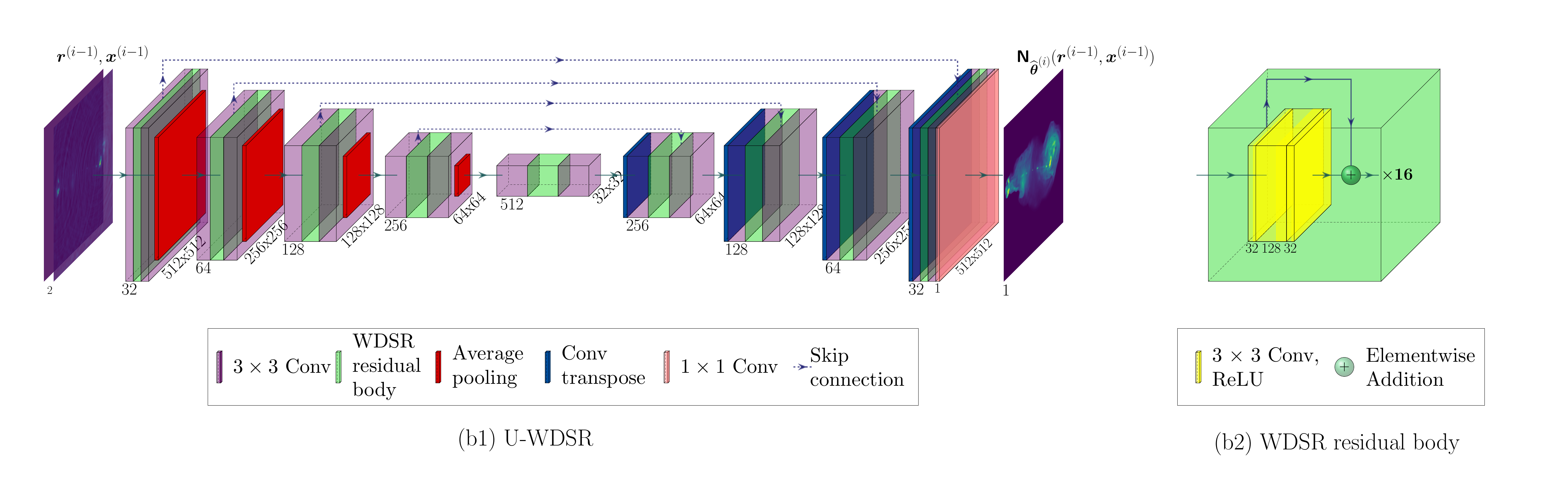} 
    \caption{R2D2 core DNN architectures. The first row panel (a) illustrates the U-Net model architecture \citep{aghabiglou2024r2d2}. The second row presents the U-WDSR model: panel (b1) shows the U-WDSR architecture and panel (b2) depicts its WDSR layer. The WDSR residual body (in green boxes) is interlaced with the convolutional layers of the U-Net. WDSR consists of 16 consecutive residual blocks. At each stage, the spatial size of feature maps is indicated at the lower center of each box. The number of channels is indicated at the outer edge of each box.}
    \label{fig:U-WDSR}    
\end{figure*}

The integrated WDSR residual body maintains an identical architecture to \citet{yu2018wide}, featuring an augmented number of blocks extended to 16. It incorporates several key features, including  (i) wide activation, (ii) dense skip connections \citep{tong2017image}, 
(iii) weight normalization, and (iv) low-rank convolutions.
Firstly, each residual block in the WDSR body expands the number of channels prior to the ReLU activation layer. This wide activation approach allows for more information to flow through the network, enabling the model to capture more intricate and detailed patterns in the data.  
Secondly, dense skip connections allow for the reuse of information by feeding feature maps from earlier layers directly into later ones. 
This design ensures that learned features remain accessible throughout the network, thus facilitating better gradient flow during training. It also enables the network to build richer and more hierarchical representations of the data.
Thirdly, weight normalization stabilizes training by reparameterizing the weight vectors. This approach improves convergence and enables the network to achieve better performance, particularly in deep models. Finally, linear low-rank convolutions balance the computational cost introduced by wide activation, reducing the dimensionality of intermediate representations while retaining critical information. Both U-Net and U-WDSR architectures are illustrated in Fig.~\ref{fig:U-WDSR}. 

In the remainder of this paper, R2D2 models taking U-Net {as the core DNN architecture ($\mathcal{A}_1$), and trained with the respective training setups $\mathcal{T}_1$ and $\mathcal{T}_2$ will be referred to as R2D2$_{\mathcal{A}_1,\mathcal{T}_1}$ and R2D2$_{\mathcal{A}_1,\mathcal{T}_2}$. Similarly, R2D2 models {taking} U-WDSR as the core architecture {($\mathcal{A}_2$)} will be referred to as R2D2$_{\mathcal{A}_2,\mathcal{T}_1}$ and R2D2$_{\mathcal{A}_2,\mathcal{T}_2}$.

\subsection{Epistemic uncertainty quantification}
Uncertainty quantification is critical for ill-posed inverse 
problems. On the one hand, incomplete data introduces aleatoric uncertainty. On the other hand, epistemic uncertainty arises from the choice of regularization models. Given the deterministic nature of R2D2, direct aleatoric uncertainty assessment is not feasible. 
In this section, we propose an ensemble averaging approach to quantify epistemic uncertainty and evaluate the robustness of R2D2 models from two perspectives. First, multiple series are trained with different random initializations of the first DNN, capturing variability arising from the training process. Second, variations in visibility-weighting schemes introduced by different Briggs parameters $\rho_\textrm{br}$ also contribute to epistemic uncertainty.

To quantify uncertainty in both cases, we define a unified evaluation approach. Specifically, we consider the concatenation of reconstructed image estimates $\widehat{\boldsymbol{X}} \in \mathbb{R}^{N \times R}$, represented as: 
\begin{equation} 
\widehat{\boldsymbol{X}} = [\widehat{\xb}_1, \dots, \widehat{\xb}_R], 
\label{eq:mean_vector} 
\end{equation}
where $r\in \{1,\dots, R\}$ indexes the reconstructed images.  For model-based epistemic uncertainty, $\widehat{\boldsymbol{X}}$ denotes the concatenation of reconstructed images resulting from different R2D2 realizations trained with distinct random initializations. For epistemic uncertainty induced by visibility weighting, $\widehat{\boldsymbol{X}}$ comprises the concatenation of images reconstructed with different Briggs parameters $\rho_\textrm{br}$.

The pixel-wise mean image $\bmu(\widehat{\boldsymbol{X}}) \in \mathbb{R}^N$ is defined as: 
\begin{equation} 
\bmu(\widehat{\boldsymbol{X}}) = \frac{1}{R}\sum_{r=1}^R \widehat{\xb}_r, \label{eq:uncertaintymean} 
\end{equation}
The relative uncertainty image, denoted as $[\bsigma/\bmu](\widehat{\boldsymbol{X}})$, represents the pixel-wise ratio of the standard deviation to the mean and is given by: 
\begin{equation}
    [\bsigma/\bmu](\widehat{\boldsymbol{X}}) = \begin{cases}
        \frac{1}{\bmu(\widehat{\boldsymbol{X}})}\sqrt{\frac{\sum_1^R({\widehat{\xb}}_{r} - \bmu(\widehat{\boldsymbol{X}}))^2}{R}} & \textrm{if } \bmu(\widehat{\boldsymbol{X}}) > 1/\widehat{a}, \\
        0 & \textrm{otherwise},
    \end{cases}
    \label{eq:uncertainty}
\end{equation}
here, $\widehat{a}>1$ represents the target dynamic range estimated as $\widehat{a}^{~-1} = \tau/\sqrt{2 \| \mathrm{Re}\{\bm{\Phi}^\dagger \bm{\Phi}\}\|_S}$  \citep{terris2022}. This formulation ensures that uncertainty is quantified only for non-zero pixels within the target dynamic range.

\subsection{Training implementation \& computational cost }
The training of {R2D2 models} was conducted using the PyTorch library in Python \citep{paszke2019pytorch}, leveraging the TorchKbNufft package \citep{muckley:20:tah} for the implementation of the measurement operator model. TorchKbNufft provides an efficient and flexible NUFFT implementation, offering options for either fast table-based interpolation or exact computation using the sparse interpolation matrix. The former was considered for RI data simulation and the computation of the residual data during training.

Training was carried out on Cirrus, a UK Tier 2 high-performance computing (HPC) facility. The utilized GPU nodes consist of two 20-core Intel Xeon Gold 6148 processors, four NVIDIA Tesla V100-SXM2-16~GB GPUs, and 384~GB of DRAM memory. 
The learning rate was fixed to $10^{-4}$, and the batch size was set to 4 for R2D2$_{\mathcal{A}_1,\mathcal{T}_2}$ and 1 for R2D2$_{\mathcal{A}_2,\mathcal{T}_2}$, {respectively,} due to GPU memory limitations. Training parameters were selected through a coarse grid search over a representative subset of the training data, to balance convergence stability, generalization performance, and computational efficiency.

Under the proposed training setup $\mathcal{T}_2$, we compare the training computational cost of the models R2D2$_{\mathcal{A}_1,\mathcal{T}_2}$ and R2D2$_{\mathcal{A}_2,\mathcal{T}_2}$, as well as the first DNN in their series as standalone end-to-end DNN models, namely U-Net, and U-WDSR, respectively. Table~\ref{table:training_cost} summarizes the key training details, including the number of iterations $I$.
The reported total computational cost in GPU hours is obtained from averaging over $R=5$ realizations of the R2D2 models.

\begin{table}
\centering
\caption{Training computation details of U-Net,
U-WDSR, R2D2$_{\mathcal{A}_1,\mathcal{T}_2}$, and R2D2$_{\mathcal{A}_2,\mathcal{T}_2}$, all trained using the training setup $\mathcal{T}_2$}
\label{table:training_cost}
\resizebox{\linewidth}{!}{%
\hspace{-2cm}
\begin{tabular}{lccccccc}
\hline
\hline
\multirow{2}{*}{Algorithm} & \multirow{2}{*}{{$I$}} & \multirow{2}{*}{$Q (\times10^6)$} & \multirow{2}{*}{$n_{\textrm{epochs}}$} & \multirow{2}{*}{$n_{\textrm{GPU}}$} & \multicolumn{3}{c}{GPU~hr}   \\
                            &                    &                                              &                            &                         & $t_{\textrm{tot.}}$ & $t_{\textrm{dat.}}$ & $t_{\textrm{tra.}}$ \\ \hline
U-Net                       & 1                  & 31                                           & 174                        & 4                       & 85.6    & 4.4     & 81.2     \\
U-WDSR                      & 1                  & 20.9                                         & 55                         & 4                       & 165.7   & 4.4     & 161.3    \\
R2D2$_{\mathcal{A}_1,\mathcal{T}_2}$                 & 25                 & 31                                           & 325                        & 4                       & 231.6   & 85.5    & 146.1   \\
R2D2$_{\mathcal{A}_2,\mathcal{T}_2}$                & 25                 & 20.9                                         & 142                        & 4                       & 420.9  & 72.6   & 348.3   \\ \hline
\end{tabular}
}

\vspace{2mm}
\parbox{\linewidth}{\footnotesize%
\textbf{Note.} The results are presented in terms of: the number of iterations ($I$), the number of learnable parameters in each network component ($Q$), and the total number of training epochs ($n_{\textrm{epochs}}$). The total computational cost is measured in GPU hours ($t_{\textrm{tot.}}$), broken down into the cost spent updating residual dirty images ($t_{\textrm{dat.}}$) and the cost used for DNN training and updating image estimates ($t_{\textrm{tra.}}$).
}
\end{table}
 
With regards to end-to-end DNN models, the training computational cost of U-WDSR is nearly twice as high as that of U-Net, mainly due to the increased complexity of the former architecture. This trend is also observed in the training of the full DNN series underpinning their corresponding R2D2 models.  
Interestingly, the computational cost of updating residual dirty images is slightly lower for the U-WDSR-based R2D2 model, R2D2$_{\mathcal{A}_2,\mathcal{T}_2}$, even though both trained the same number of DNNs. This is explained by the adopted data-pruning strategy combined with the efficiency of the advanced architecture U-WDSR. Fig.~\ref{fig:pruning} depicting the evolution of the training dataset size, R2D2$_{\mathcal{A}_1,\mathcal{T}_2}$ reaches approximately $65\%$ of its initial size by the final iteration, against almost $40\%$ for R2D2$_{\mathcal{A}_2,\mathcal{T}_2}$. 
This suggests faster convergence enabled by U-WDSR. We note that in our implementation, the sequential training of both series concluded once the evaluation metrics of the validation set converged, even though the training sets remained sufficiently large (at least $40\%$ of their original size).

\begin{figure}
    \centering
    \includegraphics[width=0.7\columnwidth]{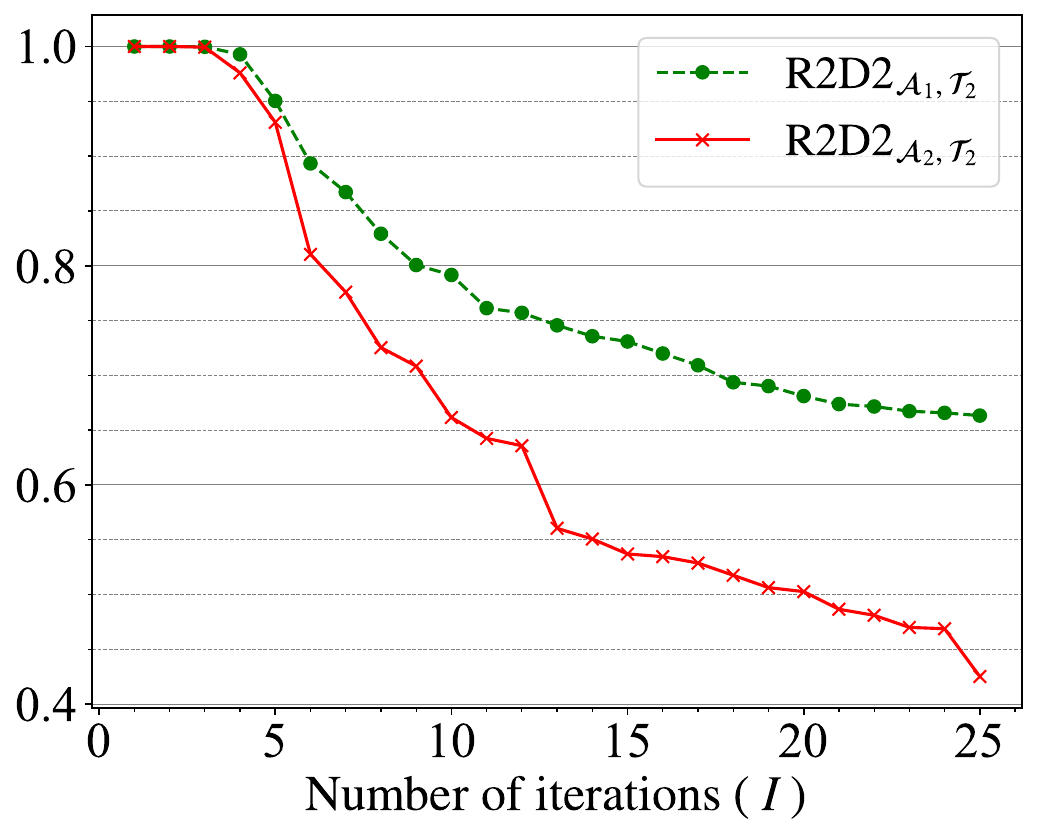}
    \caption{Evolution of the size of the training dataset throughout the iterations of R2D2$_{\mathcal{A}_1,\mathcal{T}_2}$ and R2D2$_{\mathcal{A}_2,\mathcal{T}_2}$, shown as a fraction of the size of the initial training dataset.}
    \label{fig:pruning}
\end{figure}

\section{Simulation and results} \label{SIMULATIONS&RESULTS}
This section presents a comprehensive evaluation of R2D2, focusing on its robust performance in terms of reconstruction quality and computational efficiency under various experimental setups, using VLA-specific observational settings for the formation of 512$\times$512 monochromatic intensity images. The evaluation is structured into four distinct studies. The first study compares the performance of the proposed R2D2 models to the early version. The second study benchmarks R2D2 against state-of-the-art RI algorithms. The third study quantifies R2D2’s epistemic uncertainty across its realizations. The fourth study investigates R2D2’s epistemic uncertainty under varying visibility-weighting schemes to evaluate the adaptability of R2D2 to diverse imaging conditions.

Ground-truth images used for the test dataset were derived from four real radio images, namely the giant radio galaxies 3C 353 (sourced from the NRAO Archives) and Messier~106 \citep{shimwell2022}, and the radio galaxy clusters Abell~2034 and PSZ2~G165.68+44.01 \citep{botteon2022}, following the procedure described in Section ~\ref{trainingset}.

\subsection{Benchmark algorithms \& parameter choice}\label{benchmark_algos}

R2D2 performance is studied against the RI imaging algorithms uSARA and AIRI {in BASPLib}, and multi-scale CLEAN \citep{cornwell2008multiscale} in the WSClean software \citep{offringa2014,offringa2017}. 
R2D2, AIRI, and uSARA benefit from GPU-accelerated Python implementations. 
Their core operations, including data fidelity, regularization steps for uSARA, denoising steps for AIRI, and R2D2 DNNs image reconstruction, are implemented using PyTorch. BASPLib provides four options for implementing the RI measurement operator {model}. Three of these consist in different implementations of the NUFFT, including TorchKbNufft \citep{muckley:20:tah}, FINUFFT \citep{shih2021cuFINUFFT}, and PyNUFFT \citep{lin2018python}. The fourth option leverages the PSF, which, under the assumption of a narrow field of view, enables approximating the RI mapping operator $\bm{\Phi}^\dagger \bm{\Phi}$ via a convolution with the PSF. This approach can benefit algorithms like R2D2, AIRI, and uSARA, whose iteration rules call explicitly for the dirty image and the mapping operator $\bm{\Phi}^\dagger \bm{\Phi}$ to update the residual dirty image. R2D2, uSARA and AIRI were deployed on a single GPU. As for WSClean, the software is not optimized for small-scale imaging on GPU. Therefore, it was deployed on a single CPU. Under these considerations, direct comparison of its computational performance with the GPU-accelerated algorithms is inherently unfair.

Conceptually, uSARA, AIRI and CLEAN involve free parameters that must be carefully selected. More specifically, uSARA features a parameter balancing its handcrafted regularization against data fidelity. AIRI involves a parameter controlling the choice of the DNN denoiser and the adjustment of its input to the training noise level via a scaling operation.
uSARA and AIRI parameter selection is automated using noise-driven heuristics \citep{terris2022,dabbech2022,wilber23a}. Yet, optimal results often require some tweaking around the heuristic values. In fact, in all experiments, uSARA parameter was set to twice the heuristic value. AIRI parameter was set at the heuristic for all RI data, except those simulated using ground-truth images derived from 3C 353, where $3$ times the heuristic value was considered. 
As for WSClean, multi-scale CLEAN parameters are often set to the default nominal values. However, some adjustments might be required for optimal results.  
In all experiments, auto-masking and threshold parameters of CLEAN were set to 2.0 and 0.5 times the estimated noise level, respectively. 
In contrast, R2D2 is independent of such fine-tuning requirements and is free of regularization parameters. This independence highlights a significant advantage of R2D2, enabling robust performance without the need for manual adjustments, unlike the benchmark algorithms.

\subsection{Evaluation metrics}
The reconstruction quality achieved by all algorithms is analysed through both qualitative and quantitative assessments, whereby (i) image estimates and associated residual dirty images are inspected visually, (ii) fidelity to the ground truth is evaluated using the signal-to-noise ratio (SNR) metric, computed in linear scale and logarithmic scale (logSNR), (iii) data fidelity is evaluated using the residual-to-dirty image ratio ($\textrm{RDR}$) metric, and (iv) relative uncertainty images are assessed using the mean relative uncertainty (MRU) metric, defined below.

The SNR measures the overall quality of the reconstructed image by comparing the estimate $\widehat{\xb}$ to the ground truth $\xb^\star$, and is defined as:
\begin{equation}
\textrm{SNR}(\widehat{\xb}, \xb^\star) = 20 \log_{10}\left( \frac{\|\xb^\star\|_2}{\|\xb^\star - \widehat{\xb}\|_2} \right).
\end{equation}
In high dynamic range scenarios, the logSNR metric provides a more sensitive metric for faint structures and low-intensity regions. To compute it, we first apply a logarithmic transformation to the images involved, parametrized by the target dynamic range $a$, and defined as:
\begin{equation}
\textrm{rlog}(\xb) = x_{\textrm{max}} \log_{a}\left(\frac{a}{x_{\textrm{max}}} \xb + \bm{1} \right),
\end{equation}
where $x_{\textrm{max}}$ is the peak pixel value of the image $\xb$, and $\bm{1} \in \mathbb{R}^N$ is a vector of ones. By setting $a$ to the dynamic range of the ground truth, the logSNR is computed as:
\begin{equation}
\textrm{logSNR}(\widehat{\xb}, \xb^\star) = \textrm{SNR}(\textrm{rlog}(\widehat{\xb}), \textrm{rlog}(\xb^\star)).
\end{equation}

Data fidelity is evaluated by comparing the estimated residual dirty image $\widehat{\rb}$ to the dirty image $\xb_{\textrm{d}}$. We consider the image-domain data fidelity metric, RDR, defined as:
\begin{equation}
\textrm{RDR}(\widehat{\rb}, \xb_{\textrm{d}}) = \frac{\|\widehat{\rb}\|_2}{\|\xb_{\textrm{d}}\|_2}.
\end{equation}
A lower value of $\textrm{RDR}$ indicates higher data fidelity in the image domain. 

We evaluate R2D2's robustness by examining its pixel-wise relative uncertainty images $[\bsigma/\bmu](\widehat{\boldsymbol{X}})$, obtained as per \eqref{eq:uncertainty} and report the corresponding mean relative uncertainty value denoted by MRU, which reads:
\begin{equation}
    \textrm{MRU}(\widehat{\boldsymbol{X}})  = \frac{1}{N}\sum_{n=1}^N \left([\bsigma/\bmu](\widehat{\boldsymbol{X}})\right)_n.
\end{equation}
This metric encapsulates the overall epistemic uncertainty of R2D2 models, offering insights into their stability and reliability across different R2D2 realizations and variations in $\rho_\textrm{br}$ throughout its iterations.

We also evaluate the computational performance of the imaging algorithms. This includes measuring the total number of iterations $I$, the total computational time $t_{\textrm{tot.}}$, and the average computational time per iteration for the data fidelity step $t_{\textrm{dat.}}$ and the regularization step $t_{\textrm{reg.}}$.
Since R2D2, AIRI, and uSARA were deployed on a single GPU, their computational time is reported in seconds. The same applies to WSClean, which was run on a single CPU. 

\subsection{Robust R2D2 vs. early version} \label{robust_VS_early_version}

In this study, we assess the robustness of the proposed models in comparison with the earlier model from \citet{aghabiglou2024r2d2}. In particular, we investigate the impact of (i) the choice of the core DNN architecture and (ii) the design of the training setup. 
Two experimental setups were considered. The first experimental setup, dubbed $\mathcal{E}_1$, corresponds to {the test dataset adopted in} \citet[][Table 2]{aghabiglou2024r2d2}, that is consistent with the training setup $\mathcal{T}_1$. The second experimental setup, dubbed $\mathcal{E}_2$, is fully generalized with all observational and imaging parameters uniformly randomized following the proposed training setup {$\mathcal{T}_2$.} Specifically, $\mathcal{E}_2$ is composed of $200$ inverse problems, simulated from $50$ ground-truth images of varying dynamic range for each of the four source radio images.

Reconstruction results in terms of SNR and logSNR metrics, presented in Table~\ref{tab:randomVsfix}, demonstrate that R2D2 models underpinned by the advanced U-WDSR architecture {(R2D2$_{\mathcal{A}_2,\mathcal{T}_1}$, R2D2$_{\mathcal{A}_2,\mathcal{T}_2}$)} consistently outperform the ones underpinned by U-Net (R2D2$_{\mathcal{A}_1,\mathcal{T}_1}$, R2D2$_{\mathcal{A}_1,\mathcal{T}_2}$) in both experimental setups $\mathcal{E}_1$ and $\mathcal{E}_2$. When tested on $\mathcal{E}_2$, R2D2$_{\mathcal{A}_2,\mathcal{T}_1}$ trained with fixed imaging settings still performed reliably, as opposed to R2D2$_{\mathcal{A}_1,\mathcal{T}_1}$. This highlights the robustness of the R2D2 model underpinned by the novel architecture U-WDSR and its ability to generalize beyond its training setup. When tested on {$\mathcal{E}_1$}, both R2D2$_{\mathcal{A}_1,\mathcal{T}_2}$ and R2D2$_{\mathcal{A}_2,\mathcal{T}_2}$, trained under a generalized setup, achieve a comparable performance to those trained under the more specific setup of {$\mathcal{E}_1$}. 
These findings showcase that generalizing the training setup through stochastic variations in all observational and imaging settings does not lead to suboptimal results compared to testing in a more specific setup. Moreover, they emphasize that the combination of an advanced DNN architecture, such as U-WDSR, and a diverse, well-constructed training setup significantly boosts the robustness of the R2D2 model. 

\begin{table}
\centering
\caption{Performance of the different R2D2 models under different experimental setups}
\label{tab:randomVsfix}
\begin{tabular}{lccc}
\hline
\hline
R2D2 model   & Tested on      & SNR~(dB) & logSNR~(dB) \\ 
\hline                         
R2D2$_{\mathcal{A}_1,\mathcal{T}_1}$ & \multirow{4}{*}{$\mathcal{E}_1$} & 33.7 $\pm$ 1.5   & 25.1 $\pm$ 4.9 \\ 
R2D2$_{\mathcal{A}_1,\mathcal{T}_2}$ &    & 33.2 $\pm$ 2.3 & 24.4 $\pm$ 5.3 \\
R2D2$_{\mathcal{A}_2,\mathcal{T}_1}$&    & \textbf{34.7 $\pm$ 1.6}   & \textbf{25.7 $\pm$ 4.9} \\ 
R2D2$_{\mathcal{A}_2,\mathcal{T}_2}$ &    & \textbf{34.3 $\pm$ 1.6}   & \textbf{25.6 $\pm$ 4.8} \\ 
\hline
R2D2$_{\mathcal{A}_1,\mathcal{T}_1}$     & \multirow{4}{*}{$\mathcal{E}_2$} & 20.2 $\pm$ 12.0    & 12.4 $\pm$ 12.2   \\ 
R2D2$_{\mathcal{A}_1,\mathcal{T}_2}$  &  & 30.0 $\pm$ 3.0   & 23.4 $\pm$ 4.2 \\
R2D2$_{\mathcal{A}_2,\mathcal{T}_1}$ &  & 28.6 $\pm$ 4.7     & 21.5 $\pm$ 5.6   \\ [0.3em]
R2D2$_{\mathcal{A}_2,\mathcal{T}_2}$ & & \textbf{31.2 $\pm$ 2.4}   & \textbf{24.6 $\pm$ 4.2} \\
\hline                              
\end{tabular}

\vspace{2mm}
\parbox{\linewidth}{\footnotesize%
\textbf{Note.}
Specifically, we compare the reconstruction quality (SNR and logSNR) achieved by the proposed models R2D2$_{\mathcal{A}_1,\mathcal{T}_2}$ and R2D2$_{\mathcal{A}_2,\mathcal{T}_2}$ against the earlier model R2D2$_{\mathcal{A}_1,\mathcal{T}_1}$ \citep{aghabiglou2024r2d2}. We also provide the results of the model R2D2$_{\mathcal{A}_2,\mathcal{T}_1}$. The considered experimental setups $\mathcal{E}_1$ and $\mathcal{E}_2$ are consistent with the respective training setups $\mathcal{T}_1$ and $\mathcal{T}_2$. 
All reported values represent mean $\pm$ standard deviation, calculated over 200 inverse problems. Best results are highlighted in bold.
}
\end{table}

\begin{deluxetable*}{lcccccccc}
\tabletypesize{\relsize{-2}}
\tablecaption{Evaluation of the performance of the proposed R2D2 models against benchmarking RI algorithms}
\label{tab:result_table}
\tablewidth{0pt}
\tablehead{
\colhead{Algorithm} & \colhead{SNR~(dB)} & \colhead{logSNR~(dB)} & \colhead{$\textrm{RDR}~(\times10^{-3}$)} & \colhead{$I$} & \colhead{$t_{\textrm{tot.}}$~(s)} & \colhead{$t_{\textrm{dat.}}$~(s)} & \colhead{$t_{\textrm{reg.}}$~(s)} & \colhead{$[\bm{\Phi}^\dagger \bm{\Phi}]_{\textrm{imp.}}$}
}
\startdata
CLEAN & 12.0 $\pm$ 19.3 & 9.4 $\pm$ 18.9 & 3.29 $\pm$ 2.8 & 8.4 $\pm$ 1.0 & 106.5 $\pm$ 81.6 & 11.36 $\pm$ 9.76 & 1.38 $\pm$ 0.50 & - \\
\hline
\multirow{4}{*}{uSARA} & \multirow{4}{*}{28.1 $\pm$ 3.4} & \multirow{4}{*}{20.4 $\pm$ 3.4} & \multirow{4}{*}{2.15 $\pm$ 2.7} & 1482.2 $\pm$ 586.1 & 368.8 $\pm$ 296.8 & 0.1660 $\pm$ 0.1621 & 0.0806 $\pm$ 0.0646 & TorchKbNufft \\
\cline{5-9}
& & & & 1490.7 $\pm$ 563.2 & 216.7 $\pm$ 119.3 & 0.0855 $\pm$ 0.0512 & 0.0581 $\pm$ 0.0324 & PyNUFFT \\
\cline{5-9}
& & & & 1483.1 $\pm$ 587.0 & 103.0 $\pm$ 39.52 & 0.0100 $\pm$ 0.0290 & 0.0588 $\pm$ 0.0340 & FINUFFT \\
\cline{5-9}
& & & & 1482.6 $\pm$ 586.4 & 88.03 $\pm$ 31.98 & 0.0005 $\pm$ 0.00002 & 0.0581 $\pm$ 0.0326 & PSF \\
\hline
\multirow{4}{*}{AIRI} & \multirow{4}{*}{28.3 $\pm$ 3.1} & 21.1 $\pm$ 3.8 & \multirow{4}{*}{2.24 $\pm$ 2.8} & \multirow{4}{*}{5000.0 $\pm$ 0.0} & 937.4 $\pm$ 801.8 & 0.1660 $\pm$ 0.1604 & 0.0016 $\pm$ 0.0584 & TorchKbNufft \\
\cline{3-3}
\cline{6-9}
& & \multirow{3}{*}{21.0 $\pm$ 3.8} & & & 566.5 $\pm$ 355.8 & 0.0864 $\pm$ 0.0580 & 0.0015 $\pm$ 0.0343 & PyNUFFT \\
\cline{6-9}
& & & & & 157.0 $\pm$ 36.92 & 0.0091 $\pm$ 0.0114 & 0.0013 $\pm$ 0.0216 & FINUFFT \\
\cline{6-9}
& & & & & 114.2 $\pm$ 3.450 & 0.0005 $\pm$ 0.0001 & 0.0220 $\pm$ 0.0703 & PSF \\
\hline
U-Net & 17.9 $\pm$ 3.0 & 6.8 $\pm$ 3.9 & 113.3 $\pm$ 58.9 & 1 & 0.641 $\pm$ 0.110 & - & 0.641 $\pm$ 0.110 & - \\
\hline
U-WDSR & 16.0 $\pm$ 3.6 & 6.6 $\pm$ 3.8 & 155.8 $\pm$ 81.5 & 1 & 0.662 $\pm$ 0.031 & - & 0.662 $\pm$ 0.031 & - \\
\hline
\multirow{4}{*}{R2D2$_{\mathcal{A}_1,\mathcal{T}_2}$} & \multirow{4}{*}{30.0 $\pm$ 3.0} & \multirow{4}{*}{23.4 $\pm$ 4.2} & \multirow{4}{*}{4.07 $\pm$ 9.1} & \multirow{4}{*}{18.3 $\pm$ 5.6} & 7.243 $\pm$ 4.131 & 0.2123 $\pm$ 0.1932 & 0.0462 $\pm$ 0.4572 & TorchKbNufft \\
\cline{6-9}
& & & & & 6.932 $\pm$ 3.960 & 0.1992 $\pm$ 0.1791 & 0.0197 $\pm$ 0.0686 & PyNUFFT \\
\cline{6-9}
& & & & & 3.771 $\pm$ 1.224 & 0.0356 $\pm$ 0.0685 & 0.0212 $\pm$ 0.1104 & FINUFFT \\
\cline{6-9}
& & & & & 3.342 $\pm$ 0.931 & 0.0003 $\pm$ 0.0001 & 0.0200 $\pm$ 0.0773 & PSF \\
\hline
\multirow{4}{*}{R2D2$_{\mathcal{A}_2,\mathcal{T}_2}$} & \multirow{4}{*}{31.2 $\pm$ 2.4} & \multirow{4}{*}{24.6 $\pm$ 4.2} & \multirow{4}{*}{2.22 $\pm$ 2.8} & \multirow{4}{*}{15.8 $\pm$ 5.5} & 8.831 $\pm$ 3.923 & 0.2700 $\pm$ 0.1944 & 0.1059 $\pm$ 0.3221 & TorchKbNufft \\
\cline{6-9}
& & & & & 9.023 $\pm$ 4.112 & 0.2437 $\pm$ 0.1756 & 0.0922 $\pm$ 0.1239 & PyNUFFT \\
\cline{6-9}
& & & & & 5.951 $\pm$ 2.199 & 0.0867 $\pm$ 0.0854 & 0.0878 $\pm$ 0.1055 & FINUFFT \\
\cline{6-9}
& & & & & 5.649 $\pm$ 1.911 & 0.0003 $\pm$ 0.0002 & 0.0862 $\pm$ 0.0871 & PSF \\
\enddata
\vspace{2mm}
\parbox{\linewidth}{\footnotesize%
\textbf{Note.} Reconstruction quality metrics are SNR, logSNR, and $\textrm{RDR}$. Computational performance is evaluated using the total number of iterations ($I$), the total reconstruction time ($t_{\textrm{tot.}}$), the average time per iteration for both the data fidelity step ($t_{\textrm{dat.}}$) and the regularization step ($t_{\textrm{reg.}}$). $[\bm{\Phi}^\dagger \bm{\Phi}]_{\textrm{imp.}}$ is indicating the measurement operator implementation. All reported values represent mean $\pm$ standard deviation, calculated over 200 inverse problems. For CLEAN, the reported number of iterations corresponds to the number of major cycles required for convergence. Additionally, CLEAN diverged in three test inverse problems. These cases are therefore excluded from the reported results.}
\end{deluxetable*}
\subsection {Robust R2D2 vs. benchmarking algorithms}
\label{subsec:Experiment-generic-settings}

We study the performance of the proposed R2D2 models in comparison with the benchmarking algorithms using the experimental setup $\mathcal{E}_2$ introduced in Section~\ref{robust_VS_early_version}. Numerical results of all algorithms are summarized in Table~\ref{tab:result_table}, which includes the reconstruction quality metrics as well as additional computational metrics. 
Reported values are computed as averages across all inverse problems. Additionally, results of all iterative algorithms, with the exception of CLEAN, are reported for the four different implementations of the RI mapping operator $\bm{\Phi}^\dagger \bm{\Phi}$, presented in Section~\ref{benchmark_algos}.

In terms of SNR and logSNR metrics, the results demonstrate that CLEAN and end-to-end DNN architectures (U-Net and U-WDSR) perform suboptimally. The benchmark algorithms uSARA and AIRI deliver comparable 
values, with the latter achieving marginally higher values. Interestingly, R2D2 models enable superior reconstruction quality, outperforming both uSARA and AIRI by almost $2$ to $4$~dB in both metrics. Focusing on R2D2 models, R2D2$_{\mathcal{A}_2,\mathcal{T}_2}$ yields better reconstruction results than R2D2$_{\mathcal{A}_1,\mathcal{T}_2}$, {as per the findings of Section~\ref{robust_VS_early_version}}. When examining data fidelity via the metric $\textrm{RDR}$, one can see that R2D2$_{\mathcal{A}_2,\mathcal{T}_2}$, AIRI, and uSARA are the best-performing algorithms, exhibiting comparable low values. In contrast, CLEAN delivers nearly 50$\%$ higher values, whereas R2D2$_{\mathcal{A}_1,\mathcal{T}_2}$ obtains twice as high values, on average. Finally, both end-to-end DNNs perform poorly, confirming once again the advantage of the DNN series.

With regards to the different implementations of the RI mapping operator $\bm{\Phi}^\dagger \bm{\Phi}$, R2D2, AIRI, and uSARA maintain a consistent reconstruction quality in terms of SNR and logSNR with a relative difference of the order of $10^{-4}$ on average. This is somewhat expected in the context of narrow-field small-scale imaging. However, the different implementations of $\bm{\Phi}^\dagger \bm{\Phi}$ had a significant impact on the computational efficiency of the different algorithms. Approximating the mapping operator using the PSF enabled the fastest computations of the residual dirty images (involved in the data fidelity step of the algorithms' iterative structure). The NUFFT packages exhibited varying performance, with FINUFFT being the most efficient, and TorchKbNufft the slowest of the three. Generally, both FINUFFT and the PSF-based approximation yield comparable reconstruction times for the different algorithms, whereas PyNUFFT and TorchKbNufft yield $2$ to $6$ times slower reconstructions depending on the iterative nature of the RI algorithms. While highly efficient when conducted on GPU, it is important to note that the PSF approximation can severely hamper imaging precision, particularly in wide-field imaging where the so-called $w$-effect emanating from the non-coplanarity of the radio array becomes non-negligible, or more generally, in the presence of direction-dependent effects.

In terms of computational efficiency, R2D2 models enable fast reconstructions, taking few seconds only, thanks to the combination of their limited number of iterations (hence, a few passes through the data), and the inference speed of their DNNs. This constitutes a drastic reduction in reconstruction time compared to AIRI and uSARA, both taking several minutes to converge. 
Despite AIRI’s efficient denoising steps, its larger iteration count results in longer total reconstruction times compared to uSARA. Nonetheless, thanks to their GPU implementations, both algorithms have significantly improved computational efficiency, with uSARA and AIRI being approximately 40 and 22 times faster than their CPU-based counterparts \citep{aghabiglou2024r2d2}, respectively.
R2D2 models are also faster than CLEAN. However, one must acknowledge that the considered implementation of CLEAN was not optimized for small-scale imaging on GPUs. 
Finally, both end-to-end DNN models, U-WDSR and U-Net, show an increased inference time, compared to the average execution time of DNN inference within the R2D2 series. This stems from the computational overhead incurred during DNN loading.

\subsection{Uncertainty quantification via model realizations} \label{uncertainty_realization}
We study the epistemic uncertainty of the proposed R2D2 models via ensemble averaging across different R2D2 realizations. {To this aim, we trained $R=5$ realizations for each of the models R2D2$_{\mathcal{A}_1,\mathcal{T}_2}$ and R2D2$_{\mathcal{A}_2,\mathcal{T}_2}$, and  tested them on the experimental setup $\mathcal{E}_2$ described in Section~\ref{subsec:Experiment-generic-settings}.  With $\widehat{\boldsymbol{X}}$ computed as per \eqref{eq:mean_vector} from the resulting reconstruction vectors, we analyse the pixel-wise mean {image} $\bmu(\widehat{\boldsymbol{X}})$, and the pixel-wise relative uncertainty image $[\bsigma/\bmu](\widehat{\boldsymbol{X}})$.} We also analyse the iteration-specific images $\bmu(\widehat{\boldsymbol{X}}^{(i)})$ and $[\bsigma/\bmu](\widehat{\boldsymbol{X}}^{(i)})$ for insights on the evolution of the epistemic uncertainty across the iterations of R2D2 models.

The first row of Fig.~\ref{fig:uncertainty_plots} investigates the epistemic uncertainty across R2D2 realizations by tracking its evolution over the metrics. Specifically, it presents (i) the reconstruction quality metrics, SNR and logSNR, of mean images $\bmu(\widehat{\boldsymbol{X}})$, and (ii) the mean value of the relative uncertainty image $[\bsigma/\bmu](\widehat{\boldsymbol{X}})$ denoted by MRU. 
For both R2D2 models, the mean images $\bmu(\widehat{\boldsymbol{X}})$ enable an incremental increase of both SNR and logSNR with respect to those obtained from the corresponding individual realizations. The examination of MRU reveals that although R2D2$_{\mathcal{A}_2,\mathcal{T}_2}$ exhibits higher initial uncertainty, it decreases more rapidly over iterations, ultimately achieving lower uncertainty values with a standard deviation ({i.e.}~shaded area) that is $2.5$ times lower than that of R2D2$_{\mathcal{A}_1,\mathcal{T}_2}$. This trend highlights the superior robustness of R2D2$_{\mathcal{A}_2,\mathcal{T}_2}$, achieving greater consistency in its image reconstruction as the number of iterations increases.

\begin{figure*}
    \centering
    \setlength\tabcolsep{1pt}
    \begin{tabular}{ccc}
        \includegraphics[width=0.33\linewidth]{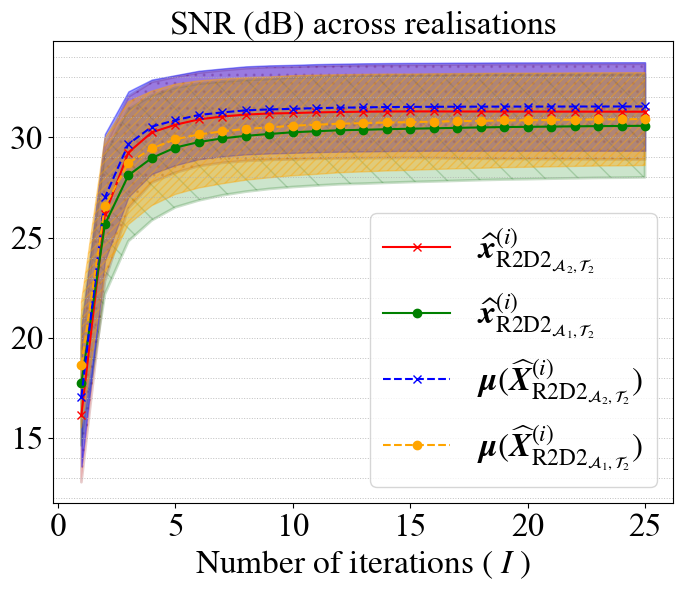} & 
        \includegraphics[width=0.33\linewidth]{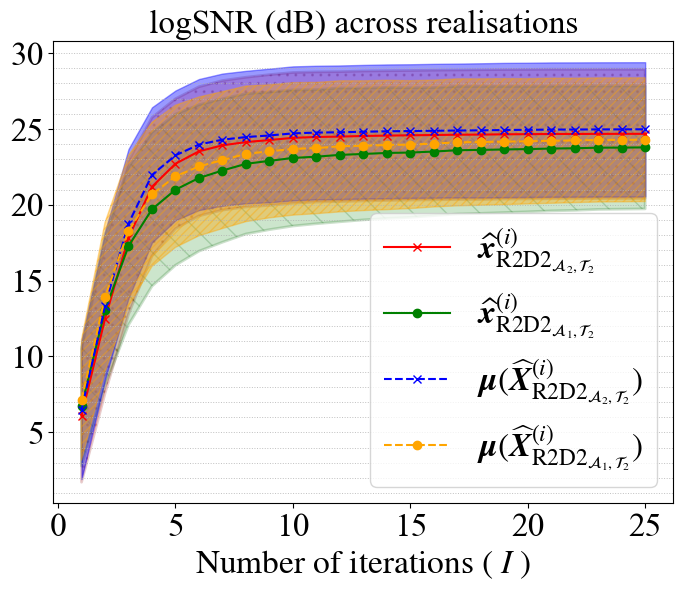} & 
        \includegraphics[width=0.33\linewidth]{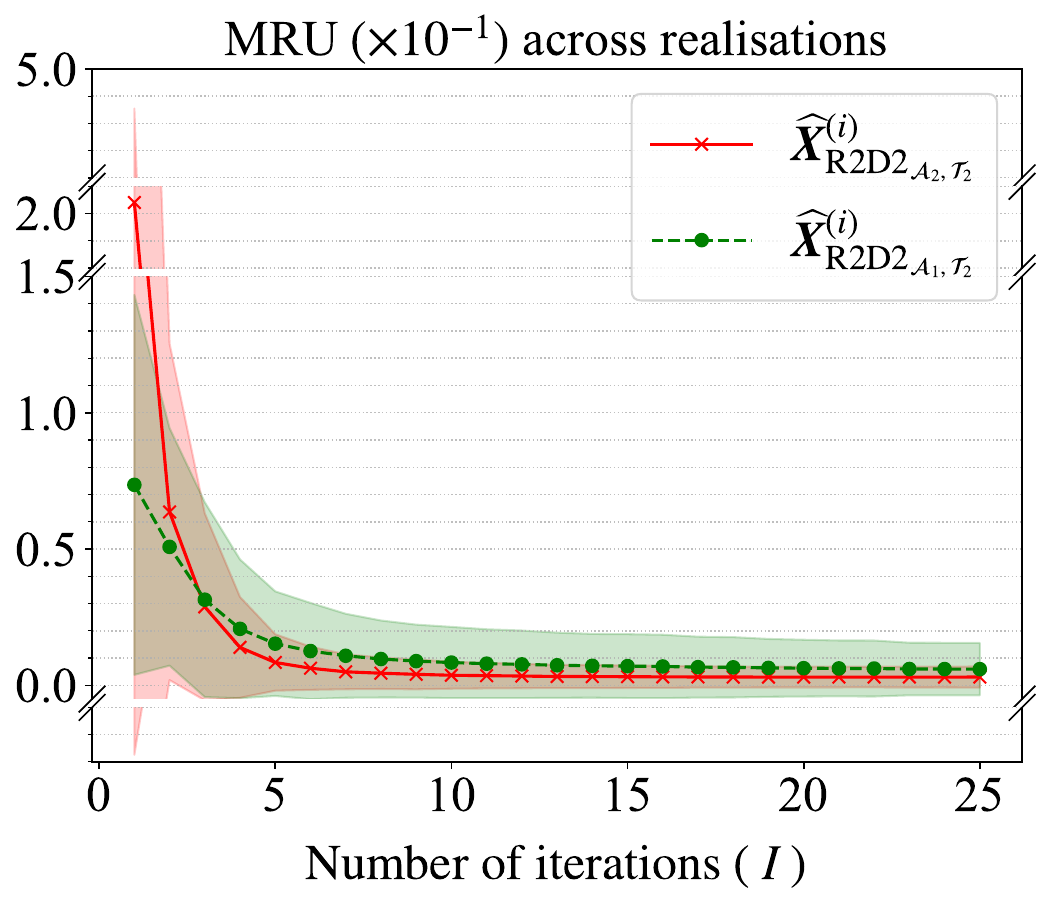} \\
        \includegraphics[width=0.33\linewidth]{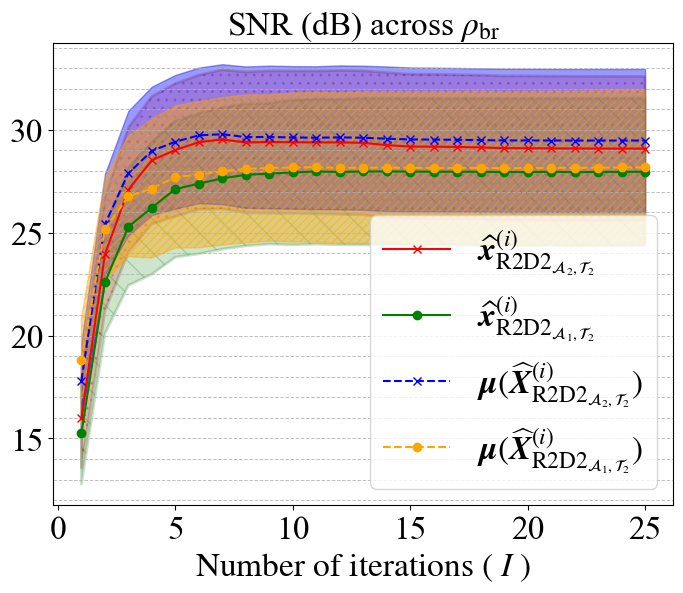} & 
        \includegraphics[width=0.33\linewidth]{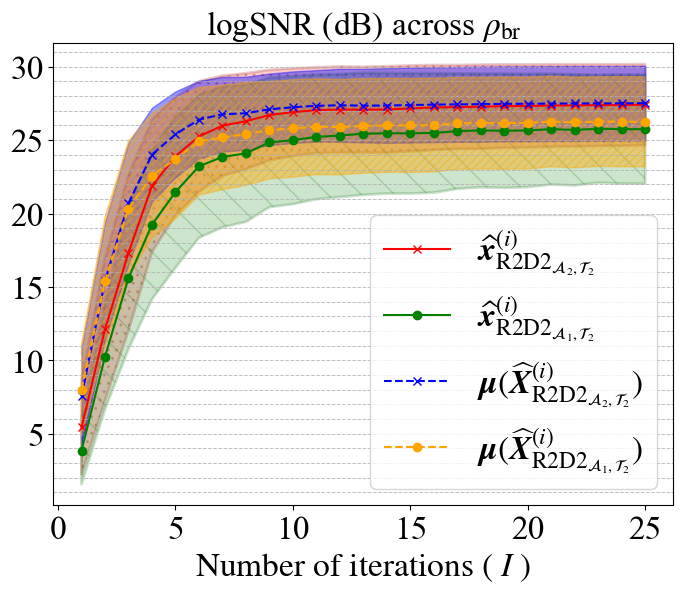} & 
        \includegraphics[width=0.33\linewidth]{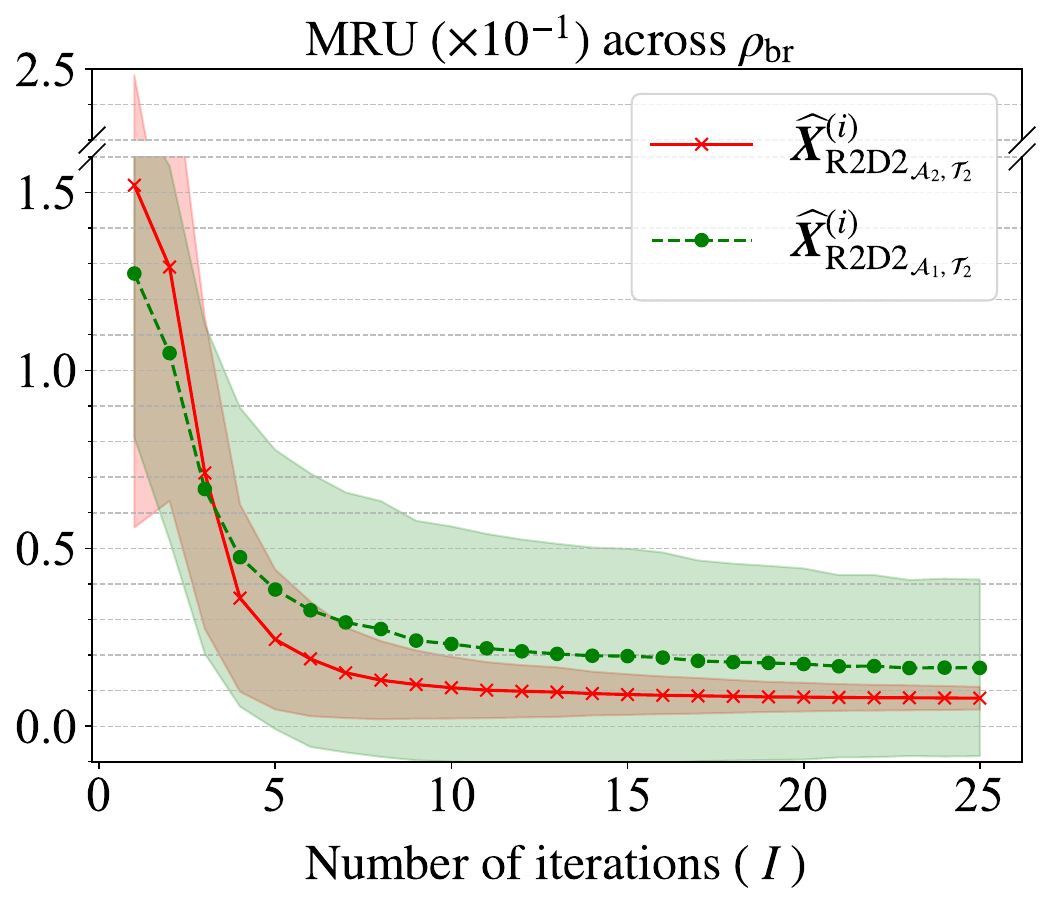}
    \end{tabular}
    \caption{Analysis of R2D2's epistemic uncertainty across R2D2 realizations (first row) and $\rho_{\textrm{br}}$ variation (second row). From left to right: evolution of the reconstruction metrics SNR and logSNR, as well as the mean of the relative uncertainty image MRU, across the iterations of R2D2 models. 
    The shaded area presents the standard deviations at each point.}
    \label{fig:uncertainty_plots}
\end{figure*}

Panel (a) of Fig.~\ref{fig:uncertainty_figs} displays the reconstruction results of a selected inverse problem simulated using the image of the radio galaxy Messier 106. This figure includes ground truth, the dirty image, the estimated images of the worst and best  
realizations of R2D2$_{\mathcal{A}_1,\mathcal{T}_2}$ and R2D2$_{\mathcal{A}_2,\mathcal{T}_2}$. It also provides their corresponding residual dirty images and relative uncertainty images $[\bsigma/\bmu](\widehat{\boldsymbol{X}})$. 
Showcasing the worst and best realizations only is motivated by the high visual consistency observed across all individual reconstructions of both R2D2$_{\mathcal{A}_1,\mathcal{T}_2}$ and R2D2$_{\mathcal{A}_2,\mathcal{T}_2}$. Even the worst-case reconstructions remain visually comparable to both the best realizations and mean images, illustrating the consistency across different model initializations.
Additionally, quantitative evaluation metrics confirm the superior performance of R2D2$_{\mathcal{A}_2,\mathcal{T}_2}$ compared to R2D2$_{\mathcal{A}_1,\mathcal{T}_2}$, which is in agreement with the findings of Section \ref{subsec:Experiment-generic-settings}. 
The inspection of the residual dirty images reveals that R2D2$_{\mathcal{A}_1,\mathcal{T}_2}$ consistently exhibits discernible structures around the pixel positions of the brightest emission as well as ringing artifacts. However, these structures are less pronounced in the images obtained by R2D2$_{\mathcal{A}_2,\mathcal{T}_2}$. 
Examination of the relative uncertainty images $[\bsigma/\bmu](\widehat{\boldsymbol{X}})$ shows reduced uncertainty enabled by R2D2$_{\mathcal{A}_2,\mathcal{T}_2}$.   
These findings highlight the enhanced robustness and precision of R2D2$_{\mathcal{A}_2,\mathcal{T}_2}$ over R2D2$_{\mathcal{A}_1,\mathcal{T}_2}$.
\begin{figure*} 
    \centering
    \setlength\tabcolsep{2pt}
    \renewcommand{\arraystretch}{0.0}
    {\fontsize{5.8}{2} \selectfont
    \begin{tabular}{lc}
            \hspace{-1.6cm}
        (a) Epistemic uncertainty over R2D2 realizations & \hspace{1.8cm}
        (b) Epistemic uncertainty over $\rho_{\textrm{br}}$ variation \\   
    \end{tabular}
    \begin{tabular}{cclcccl} 
        \begin{tikzpicture}
            \node[inner sep=0pt] at (0,0) {\includegraphics[width=0.18\linewidth]{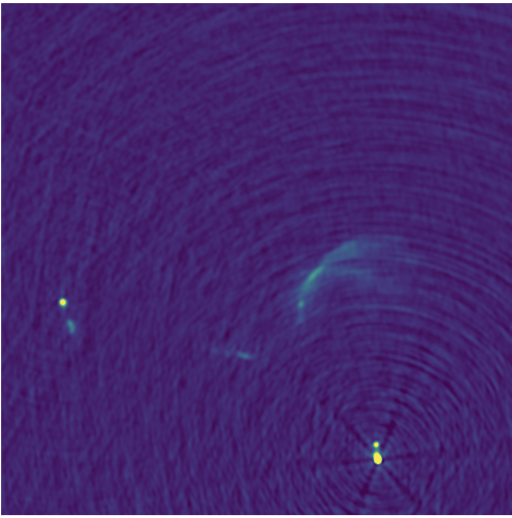}};
            \node at (-0.6,1.4) {\textcolor{white}{{Dirty image $\xb_{\textrm{d}}$}}};
        \end{tikzpicture} 
        &
        \begin{tikzpicture}
            \node[inner sep=0pt] at (0,0) {\includegraphics[width=0.18\linewidth]{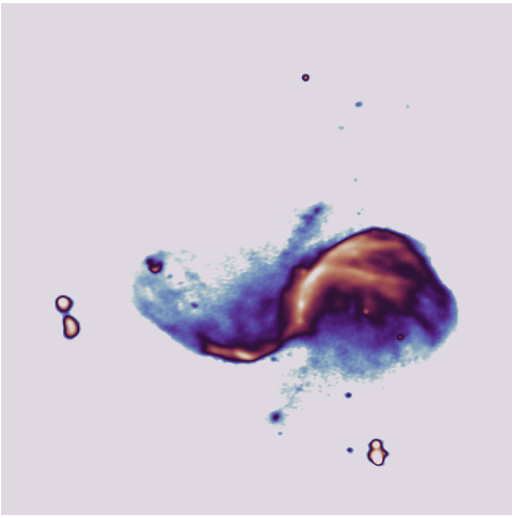}};
            \node at (-0.5,1.4) {\textcolor{black}{{Ground truth $\xb^\star$}}};
        \end{tikzpicture}
        & \hspace{-0.4cm}
        \begin{tikzpicture}
            \node[inner sep=0pt] at (0,0) {\includegraphics[width=0.024\linewidth]{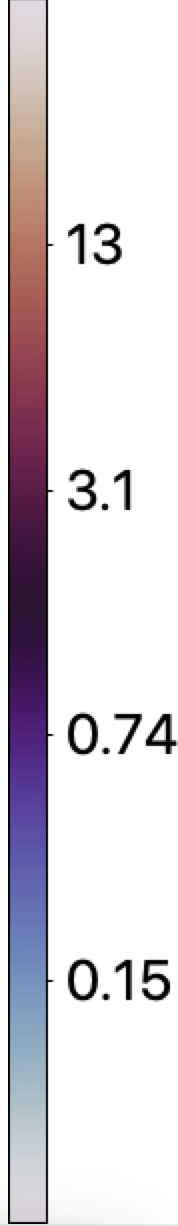}};
            \node at (0,1.65) {\textcolor{black}{$\times 10^{-3}$}};
        \end{tikzpicture} 
        & 
        &
        \begin{tikzpicture}
            \node[inner sep=0pt] at (0,0) {\includegraphics[width=0.18\linewidth]{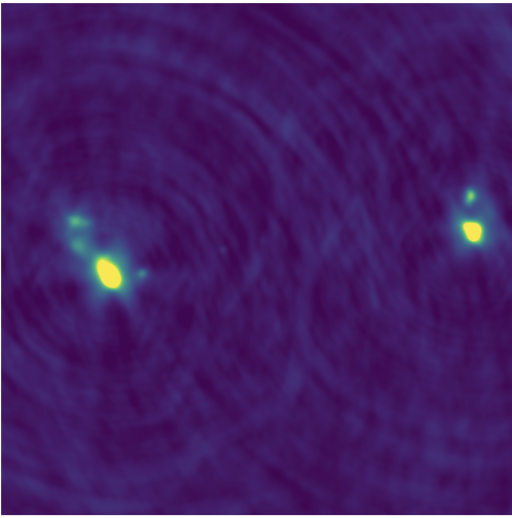}};
            \node at (0.0,1.4) {\textcolor{white}{{Dirty image $\xb_{\textrm{d}}\Rightarrow \rho_{\textrm{br}}=1$}}};
        \end{tikzpicture} 
        &
        \begin{tikzpicture}
            \node[inner sep=0pt] at (0,0) {\includegraphics[width=0.18\linewidth]{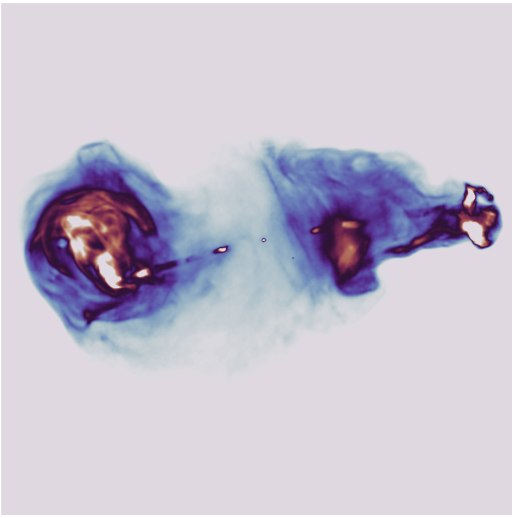}};
            \node at (-0.5,1.4) {\textcolor{black}{{Ground truth $\xb^\star$}}};
        \end{tikzpicture}        
        & \hspace{-0.4cm}
        \begin{tikzpicture}
            \node[inner sep=0pt] at (0,0) {\includegraphics[width=0.024\linewidth]{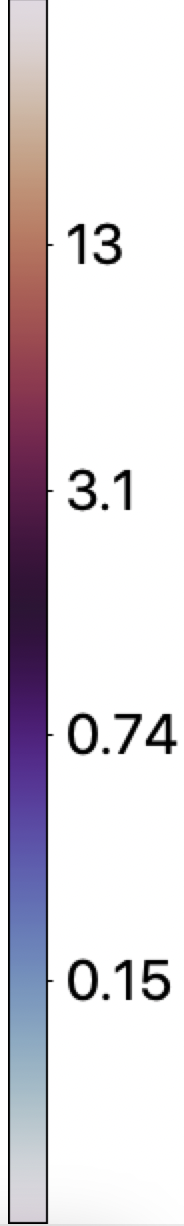}};
            \node at (0,1.65) {\textcolor{black}{$\times 10^{-3}$}};
        \end{tikzpicture} 
        \\

        \begin{tikzpicture}
            \node[inner sep=0pt] at (0,0) {\includegraphics[width=0.18\linewidth]{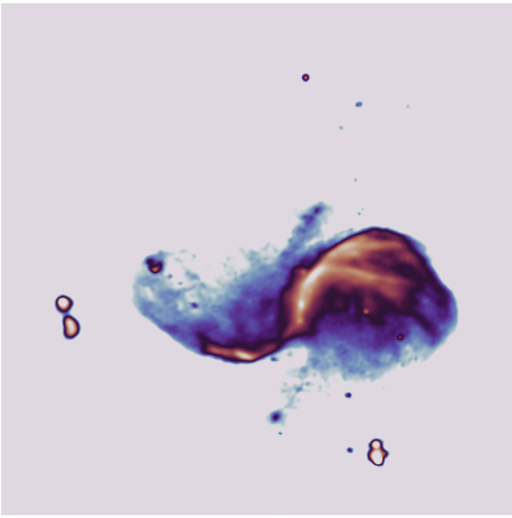}};
            \node at (0.0,1.35) {\textcolor{black}{{$\widehat{\xb}_{\textrm{R2D2}_{\mathcal{A}_1,\mathcal{T}_2}} \Rightarrow$  Worst Realis.}}};
            \node at (0.0,-1.45) {\textcolor{black}{{\tiny(SNR: 33.4, logSNR: 28.2) dB}}};
        \end{tikzpicture} 
        & 
        \begin{tikzpicture}
            \node[inner sep=0pt] at (0,0) {\includegraphics[width=0.18\linewidth]{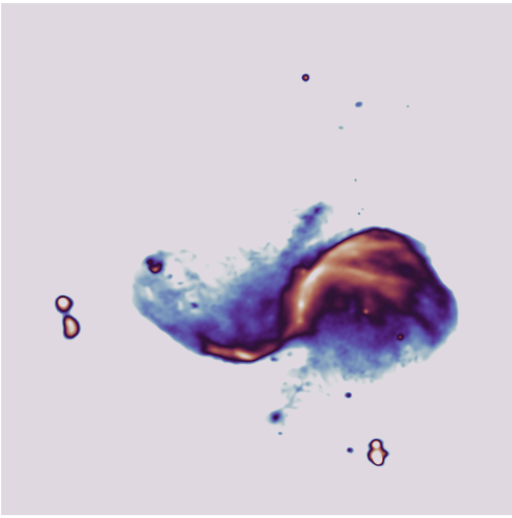}};
            \node at (0.0,1.35) {\textcolor{black}{{$\widehat{\xb}_{\textrm{R2D2}_{\mathcal{A}_2,\mathcal{T}_2}}\Rightarrow$  Worst Realis.}}};
            \node at (0.0,-1.45) {\textcolor{black}{{\tiny(SNR: 34.8, logSNR: 28.8 dB)}}};
        \end{tikzpicture}
        &
        \hspace{-0.4cm}
        \begin{tikzpicture}
            \node[inner sep=0pt] at (0,0) {\includegraphics[width=0.023\linewidth]{figs/uncertainty/GT_colourbar_vertical_1e3.png}};
            \node at (0,1.65) {\textcolor{black}{$\times 10^{-3}$}};
        \end{tikzpicture}
        & 
        
        &
        \begin{tikzpicture}
            \node[inner sep=0pt] at (0,0) {\includegraphics[width=0.18\linewidth]{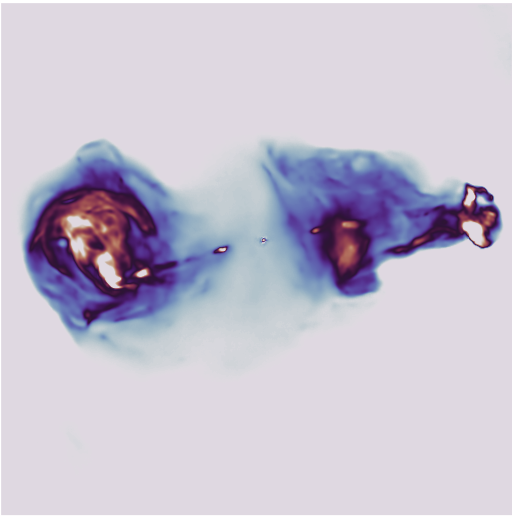}};
            \node at (-0.1,1.4) {\textcolor{black}{{$\widehat{\xb}_{\textrm{R2D2}_{\mathcal{A}_1,\mathcal{T}_2}}\Rightarrow \rho_{\textrm{br}}=-1$}}};
            \node at (0.0,-1.45) {\textcolor{black}{{\tiny(SNR: 24.0, logSNR: 25.0) dB}}};
        \end{tikzpicture} 
        & 
        \begin{tikzpicture}
            \node[inner sep=0pt] at (0,0) {\includegraphics[width=0.18\linewidth]{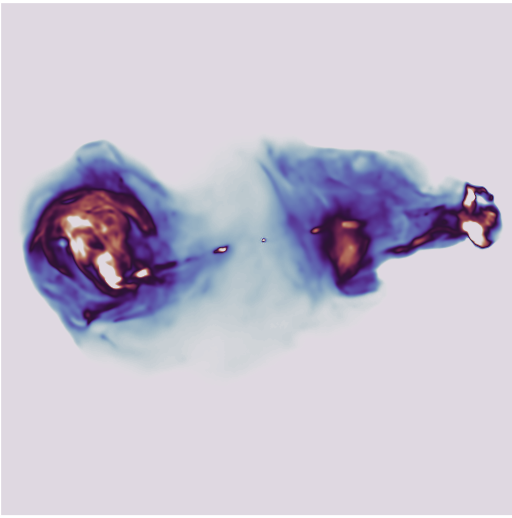}};
            \node at (-0.1,1.4) {\textcolor{black}{{$\widehat{\xb}_{\textrm{R2D2}_{\mathcal{A}_2,\mathcal{T}_2}} \Rightarrow \rho_{\textrm{br}}=-1$}}};
            \node at (0.0,-1.45) {\textcolor{black}{{\tiny(SNR: 28.1, logSNR: 26.8) dB}}};
        \end{tikzpicture} 
        &
        \hspace{-0.4cm}
        \begin{tikzpicture}
            \node[inner sep=0pt] at (0,0) {\includegraphics[width=0.024\linewidth]{figs/BR_study/GT_colourbar_vertical_1e3.png}};
            \node at (0,1.65) {\textcolor{black}{$\times 10^{-3}$}};
        \end{tikzpicture}
        \\ 
        
        \begin{tikzpicture}
            \node[inner sep=0pt] at (0,0) {\includegraphics[width=0.18\linewidth]{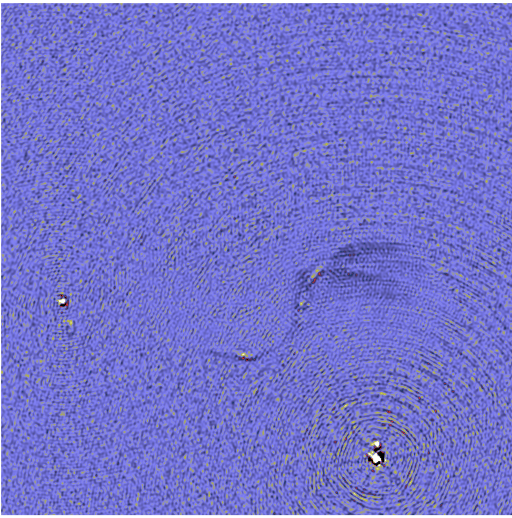}};
            \node at (0.0,1.35) {\textcolor{white}{{$\widehat{\rb}_{\textrm{R2D2}_{\mathcal{A}_1,\mathcal{T}_2}} \Rightarrow$ Worst Realis.}}};
            \node at (-0.45,-1.4) {\textcolor{white}{{$\textrm{RDR}$: 7.46 $\times 10^{-4}$}}}; 
        \end{tikzpicture} 
        & 
        \begin{tikzpicture}
            \node[inner sep=0pt] at (0,0) {\includegraphics[width=0.18\linewidth]{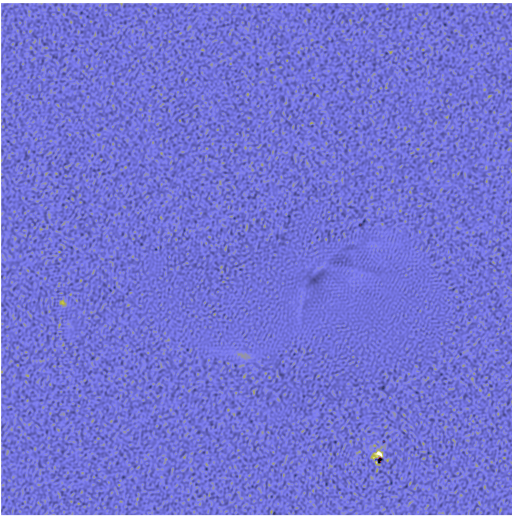}};
            \node at (0.0,1.35) {\textcolor{white}{{$\widehat{\rb}_{\textrm{R2D2}_{\mathcal{A}_2,\mathcal{T}_2}}\Rightarrow$ Worst Realis.}}};
            \node at (-0.45,-1.4) {\textcolor{white}{{$\textrm{RDR}$: 4.51 $\times 10^{-4}$}}};
        \end{tikzpicture} 
        & 
        \hspace{-0.4cm}
        \begin{tikzpicture}
            \node[inner sep=0pt] at (0,0) {\includegraphics[width=0.023\linewidth]{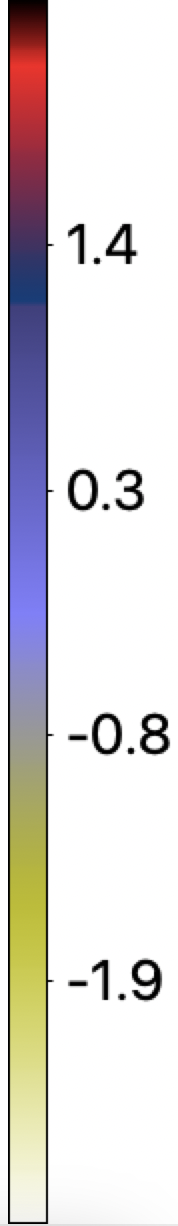}};
            \node at (0,1.65) {\textcolor{black}{$\times 10^{-4}$}};
        \end{tikzpicture}
        &
        & 
        \begin{tikzpicture}
            \node[inner sep=0pt] at (0,0) {\includegraphics[width=0.18\linewidth]{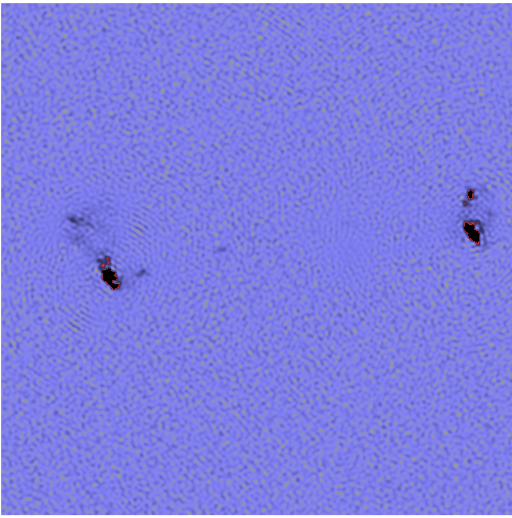}};
            \node at (-0.1,1.4) {\textcolor{white}{{$\widehat{\rb}_{\textrm{R2D2}_{\mathcal{A}_1,\mathcal{T}_2}}\Rightarrow \rho_{\textrm{br}}=-1$}}};
            \node at (-0.45,-1.4) {\textcolor{white}{{$\textrm{RDR}$: 1.12 $\times 10^{-3}$}}};
        \end{tikzpicture} 
        & 
        \begin{tikzpicture}
            \node[inner sep=0pt] at (0,0) {\includegraphics[width=0.18\linewidth]{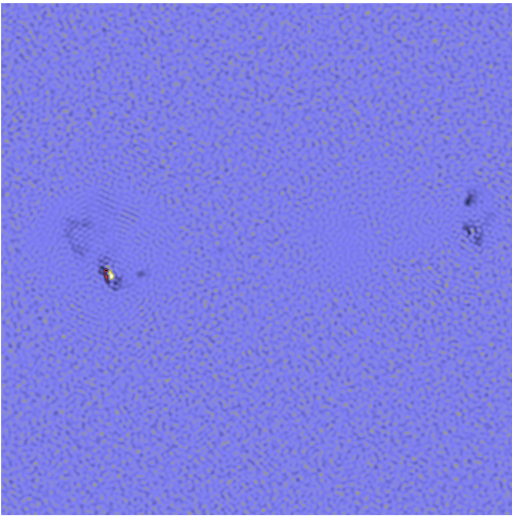}};
            \node at (-0.1,1.4) {\textcolor{white}{{$\widehat{\rb}_{\textrm{R2D2}_{\mathcal{A}_2,\mathcal{T}_2}}\Rightarrow \rho_{\textrm{br}}=-1$}}};
            \node at (-0.45,-1.4) {\textcolor{white}{{$\textrm{RDR}$: 9.27 $\times 10^{-4}$}}};
        \end{tikzpicture} 
        &
        \hspace{-0.4cm}
        \begin{tikzpicture}
            \node[inner sep=0pt] at (0,0) {\includegraphics[width=0.021\linewidth]{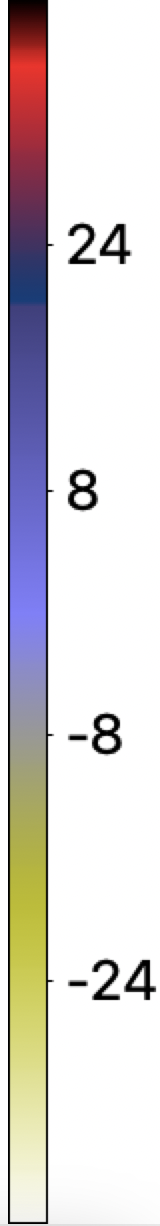}};
            \node at (0,1.65) {\textcolor{black}{$\times 10^{-4}$}};
        \end{tikzpicture}
        \\ 
        \begin{tikzpicture}
            \node[inner sep=0pt] at (0,0) {\includegraphics[width=0.18\linewidth]{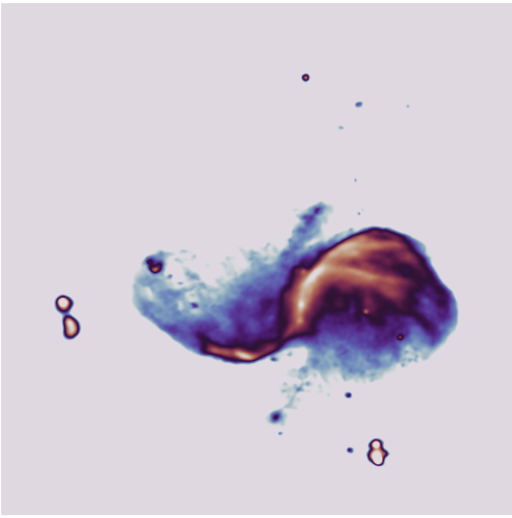}};
            \node at (0.0,1.35) {\textcolor{black}{{$\widehat{\xb}_{\textrm{R2D2}_{\mathcal{A}_1,\mathcal{T}_2}}\Rightarrow$ Best Realis.}}};
            \node at (0.0,-1.45) {\textcolor{black}{{\tiny(SNR: 33.9, logSNR: 28.2) dB}}};
        \end{tikzpicture} 
        & 
        \begin{tikzpicture}
            \node[inner sep=0pt] at (0,0) {\includegraphics[width=0.18\linewidth]{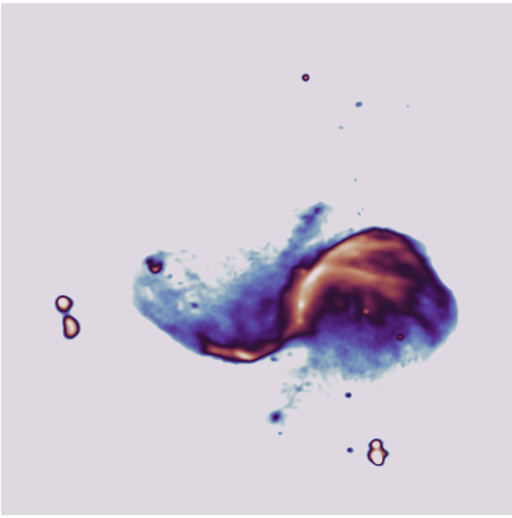}};
            \node at (0.0,1.35) {\textcolor{black}{{$\widehat{\xb}_{\textrm{R2D2}_{\mathcal{A}_2,\mathcal{T}_2}}\Rightarrow$ Best Realis.}}};
            \node at (0.0,-1.45) {\textcolor{black}{{\tiny(SNR: 35.5, logSNR: 29.2) dB}}};
        \end{tikzpicture} 
        &
        \hspace{-0.4cm}
        \begin{tikzpicture}
            \node[inner sep=0pt] at (0,0) {\includegraphics[width=0.022\linewidth]{figs/uncertainty/GT_colourbar_vertical_1e3.png}};
            \node at (0,1.65) {\textcolor{black}{$\times 10^{-3}$}};
        \end{tikzpicture}
        &
        & 
        \begin{tikzpicture}
            \node[inner sep=0pt] at (0,0) {\includegraphics[width=0.18\linewidth]{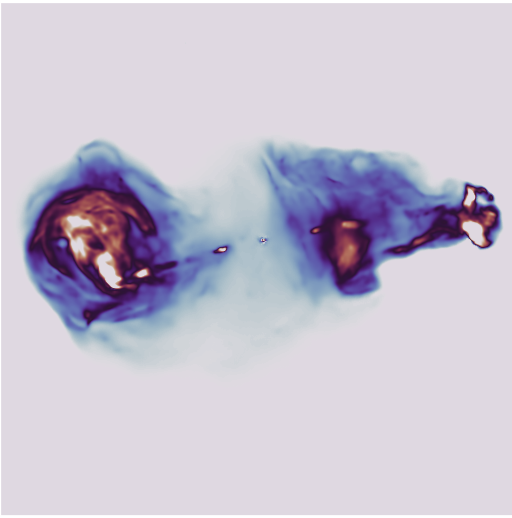}};
            \node at (-0.1,1.4) {\textcolor{black}{{$\widehat{\xb}_{\textrm{R2D2}_{\mathcal{A}_1,\mathcal{T}_2}}\Rightarrow \rho_{\textrm{br}}=1$}}};
            \node at (0.0,-1.45) {\textcolor{black}{{\tiny(SNR: 25.6, logSNR: 25.9) dB}}};
        \end{tikzpicture} 
        & 
        \begin{tikzpicture}
            \node[inner sep=0pt] at (0,0) {\includegraphics[width=0.18\linewidth]{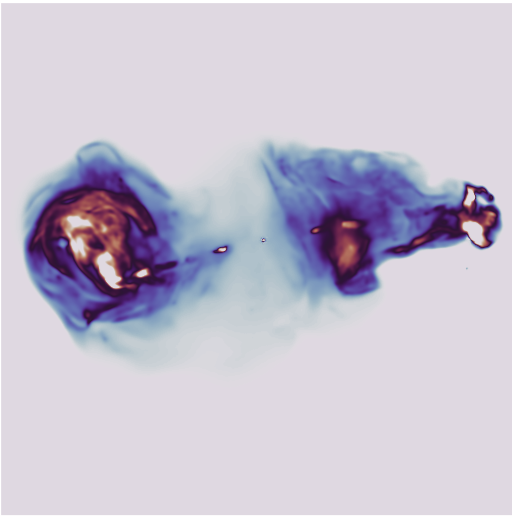}};
            \node at (-0.1,1.4) {\textcolor{black}{{$\widehat{\xb}_{\textrm{R2D2}_{\mathcal{A}_2,\mathcal{T}_2}}\Rightarrow \rho_{\textrm{br}}=1$}}};
            \node at (0.0,-1.45) {\textcolor{black}{{\tiny(SNR: 27.7, logSNR: 29.6) dB}}};
        \end{tikzpicture}  
        &
        \hspace{-0.4cm}
        \begin{tikzpicture}
            \node[inner sep=0pt] at (0,0) {\includegraphics[width=0.024\linewidth]{figs/BR_study/GT_colourbar_vertical_1e3.png}};
            \node at (0,1.65) {\textcolor{black}{$\times 10^{-3}$}};
        \end{tikzpicture}
        \\ 
         \begin{tikzpicture}
            \node[inner sep=0pt] at (0,0) {\includegraphics[width=0.18\linewidth]{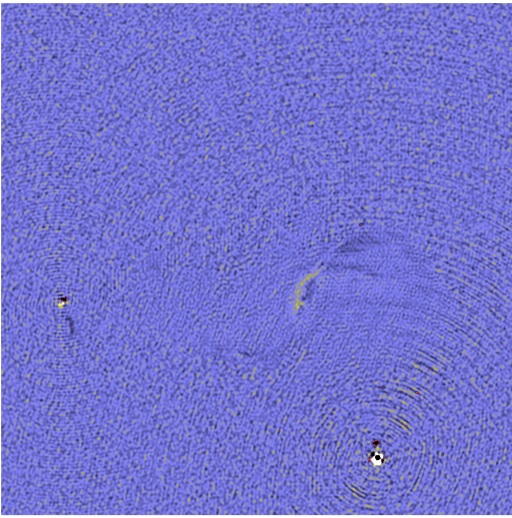}};
            \node at (0.0,1.35) {\textcolor{white}{{$\widehat{\rb}_{\textrm{R2D2}_{\mathcal{A}_1,\mathcal{T}_2}}\Rightarrow$ Best Realis.}}};
            \node at (-0.45,-1.45) {\textcolor{white}{{$\textrm{RDR}$: 6.16 $\times 10^{-4}$}}};
        \end{tikzpicture} 
        & 
        \begin{tikzpicture}
            \node[inner sep=0pt] at (0,0) {\includegraphics[width=0.18\linewidth]{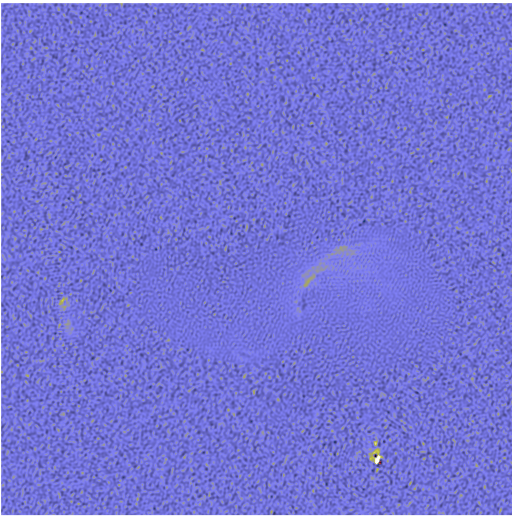}};
            \node at (-0.45,-1.45) {\textcolor{white}{{$\textrm{RDR}$: 4.59 $\times 10^{-4}$}}};
            \node at (0.0,1.35) {\textcolor{white}{{$\widehat{\rb}_{\textrm{R2D2}_{\mathcal{A}_2,\mathcal{T}_2}}\Rightarrow$ Best Realis.}}};
        \end{tikzpicture} &
        \hspace{-0.4cm}
        \begin{tikzpicture}
            \node[inner sep=0pt] at (0,0) {\includegraphics[width=0.022\linewidth]{figs/uncertainty/res_colorbar_vertical_1e4.png}};
            \node at (0,1.65) {\textcolor{black}{$\times 10^{-4}$}};
        \end{tikzpicture}
        & 
        & 
         \begin{tikzpicture}
            \node[inner sep=0pt] at (0,0) {\includegraphics[width=0.18\linewidth]{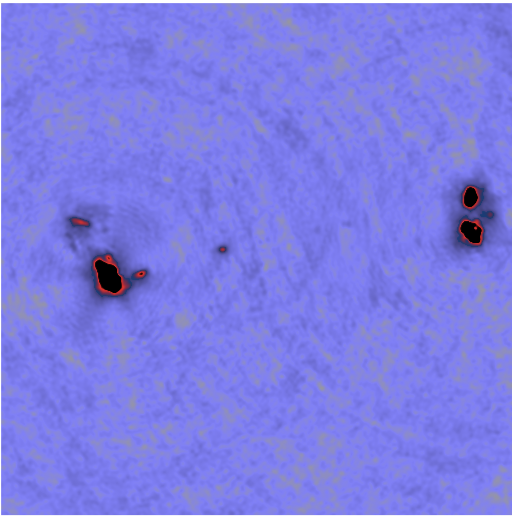}};
            \node at (-0.1,1.4) {\textcolor{white}{{$\widehat{\rb}_{\textrm{R2D2}_{\mathcal{A}_1,\mathcal{T}_2}}\Rightarrow \rho_{\textrm{br}}=1$}}};
            \node at (-0.45,-1.45) {\textcolor{white}{{$\textrm{RDR}$: 5.72 $\times 10^{-4}$}}};
        \end{tikzpicture} 
        & 
        \begin{tikzpicture}
            \node[inner sep=0pt] at (0,0) {\includegraphics[width=0.18\linewidth]{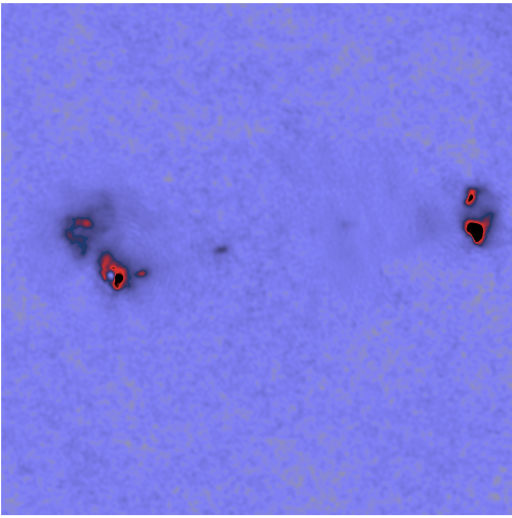}};
            \node at (-0.1,1.4) {\textcolor{white}{{$\widehat{\rb}_{\textrm{R2D2}_{\mathcal{A}_2,\mathcal{T}_2}}\Rightarrow \rho_{\textrm{br}}=1$}}};
            \node at (-0.45,-1.45) {\textcolor{white}{{$\textrm{RDR}$: 3.24 $\times 10^{-4}$}}};
        \end{tikzpicture} 
        &
        \hspace{-0.4cm}
        \begin{tikzpicture}
            \node[inner sep=0pt] at (0,0) {\includegraphics[width=0.021\linewidth]{figs/BR_study/res_colorbar_vertical_1e4_br-1.png}};
            \node at (0,1.65) {\textcolor{black}{$\times 10^{-4}$}};
        \end{tikzpicture}
        \\ 
        \begin{tikzpicture}
            \node[inner sep=0pt] at (0,0) {\includegraphics[width=0.18\linewidth]{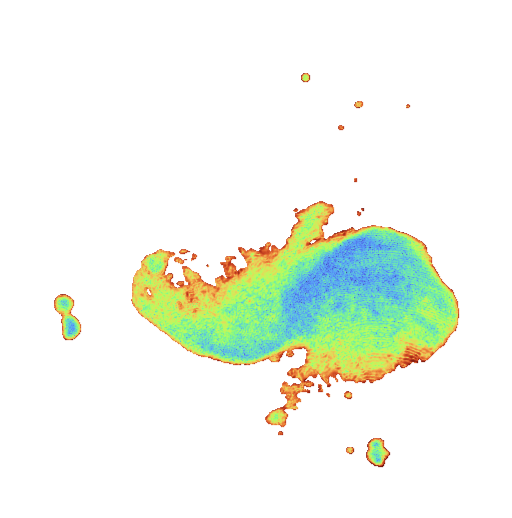}};
            \node at (-0.45,1.35) {\textcolor{black}{{$[\bsigma/\bmu](\widehat{\boldsymbol{X}}_{\textrm{R2D2}_{\mathcal{A}_1,\mathcal{T}_2}})$}}};
            \node at (-0.45,-1.5) {\textcolor{black}{MRU$:2.69 \times 10^{-3}$}};
        \end{tikzpicture}
        & 
        \begin{tikzpicture}
            \node[inner sep=0pt] at (0,0) {\includegraphics[width=0.18\linewidth]{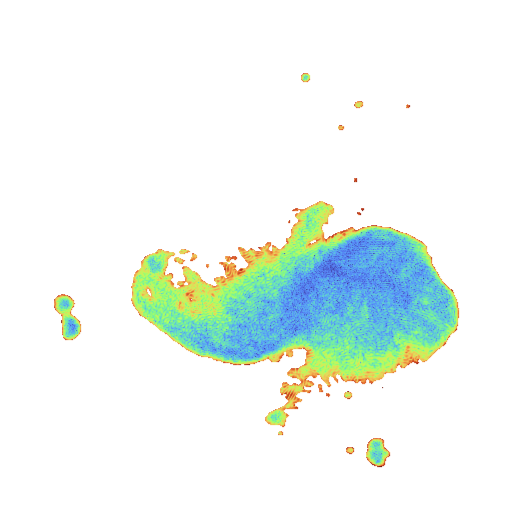}};
            \node at (-0.45,1.35) {\textcolor{black}{{$[\bsigma/\bmu](\widehat{\boldsymbol{X}}_{\textrm{R2D2}_{\mathcal{A}_2,\mathcal{T}_2}})$}}};
            \node at (-0.45,-1.5) {\textcolor{black}{MRU$:1.37 \times 10^{-3}$}};
        \end{tikzpicture} 
        &
        \hspace{-0.4cm}
        \begin{tikzpicture}
            \node[inner sep=0pt] at (0,0) {\includegraphics[width=0.023\linewidth]{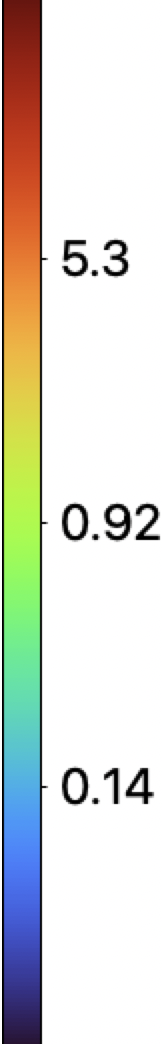}};
            \node at (0,1.65) {\textcolor{black}{$\times 10^{-2}$}};
        \end{tikzpicture}
        & 
        & 
        \begin{tikzpicture}
            \node[inner sep=0pt] at (0,0) {\includegraphics[width=0.18\linewidth]{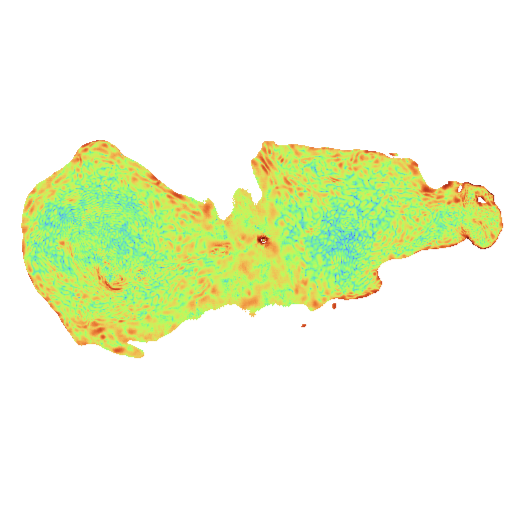}};
            \node at (-0.45,1.35) {\textcolor{black}{{$[\bsigma/\bmu](\widehat{\boldsymbol{X}}_{\textrm{R2D2}_{\mathcal{A}_1,\mathcal{T}_2}})$}}};
            \node at (-0.45,-1.5) {\textcolor{black}{MRU$:6.13 \times 10^{-3}$}};
        \end{tikzpicture} 
        & 
        \begin{tikzpicture}
            \node[inner sep=0pt] at (0,0) {\includegraphics[width=0.18\linewidth]{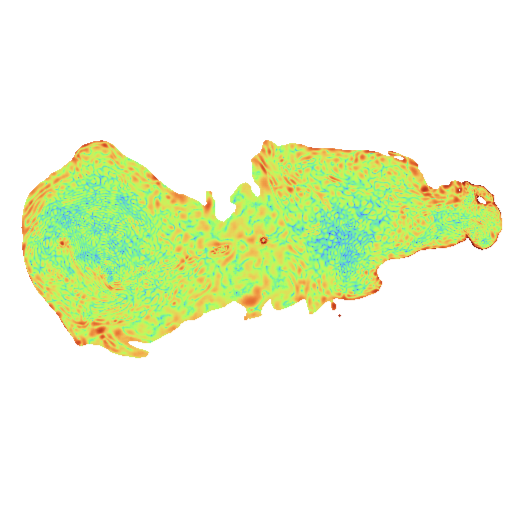}};
            \node at (-0.45,1.35) {\textcolor{black}{{$[\bsigma/\bmu](\widehat{\boldsymbol{X}}_{\textrm{R2D2}_{\mathcal{A}_2,\mathcal{T}_2}})$}}};
            \node at (-0.45,-1.5) {\textcolor{black}{ MRU$:5.93 \times 10^{-3}$}};
        \end{tikzpicture}  &
        \hspace{-0.4cm}
        \begin{tikzpicture}
            \node[inner sep=0pt] at (0,0) {\includegraphics[width=0.023\linewidth]{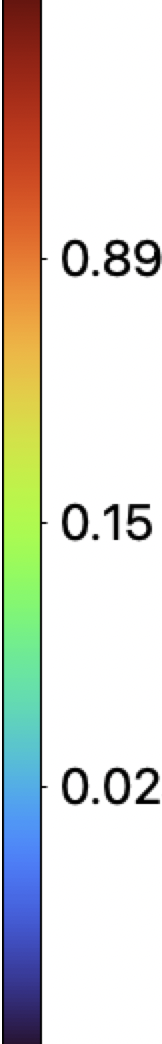}};
            \node at (0,1.65) {\textcolor{black}{$\times 10^{-1}$}};
        \end{tikzpicture}
        \\ 
        
    \end{tabular}}
    \caption{Illustration of R2D2's joint image estimation and uncertainty quantification functionality on selected RI simulations. Panel (a) focuses on epistemic uncertainty across R2D2 realizations utilizing an image of Messier 106. Panel (b) focuses on epistemic uncertainty across variations of the parameter of Briggs weighting ($\rho_\textrm{br}$) utilizing an image of 3C 353. The first row in both panels displays the dirty image (left) and ground-truth image (right). In Panel (a) (resp. panel (b)), second and fourth rows show the respective estimated images for worst and best realizations of R2D2$_{\mathcal{A}_1,\mathcal{T}_2}$ (left) and R2D2$_{\mathcal{A}_2,\mathcal{T}_2}$ (right) (resp. estimated images with $\rho_\textrm{br}=-1$ and $\rho_\textrm{br}=1$). Third and fifth rows in both panels show the corresponding residual dirty images. The sixth row displays the relative uncertainty image $[\bsigma/\bmu](\widehat{\boldsymbol{X}})$. 
    Metrics are reported inside the associated images.
    }
    
    \label{fig:uncertainty_figs}
\end{figure*}

\subsection{Uncertainty quantification via visibility weighting} \label{uncertainty_briggs}

In this study, we evaluate the epistemic uncertainty quantification of the proposed R2D2 models by performing ensemble averaging over reconstructions obtained with different values of Briggs parameter $\rho_\textrm{br}$. 
We introduce the experimental setup $\mathcal{E}_3$, comprising 1000 inverse problems with ground-truth images obtained from the image of 3C 353, with varying dynamic ranges and observational settings consistent with the training setting described in Section~\ref{trainingset}.
This transition from $\mathcal{E}_2$ (which consisted of 200 inverse problems) was necessary, as the smaller experimental setup led to instability in the results. Increasing the number of inverse problems ensures a more comprehensive and reliable evaluation.
For each inverse problem, we generate five dirty images by back-projecting the simulated RI data to the image domain using Briggs weighting and considering different values of the Briggs parameter $\rho_\textrm{br} \in \{1, 0.5, 0, -0.5, -1\}$. 

The second row of Fig.~\ref{fig:uncertainty_plots} examines the epistemic uncertainty introduced by varying the $\rho_\textrm{br}$. It tracks the evolution of the reconstruction quality metrics SNR and logSNR, and the uncertainty evaluation metric MRU throughout the iterations.
At each iteration, mean images are computed by averaging over the 1000 inverse problems. In contrast, the metrics for ${\widehat{\xb}_{\textrm{R2D2}_{\mathcal{A}_1,\mathcal{T}_2}}}$ and ${\widehat{\xb}_{\textrm{R2D2}_{\mathcal{A}_2,\mathcal{T}_2}}}$ are averaged across all 5000 inverse problems, encompassing all chosen values of  $\rho_\textrm{br}$.
Consistent with the behavior observed in Section \ref{uncertainty_realization}, the mean images $\bmu(\widehat{\boldsymbol{X}})$ for both R2D2 models show a slight improvement in both SNR and logSNR compared to those obtained from the corresponding individual reconstructions. Furthermore, the metric MRU indicates a higher initial uncertainty for R2D2$_{\mathcal{A}_2,\mathcal{T}_2}$ across variations in $\rho_\textrm{br}$ but ultimately converges to a more robust result compared to R2D2$_{\mathcal{A}_1,\mathcal{T}_2}$. Specifically, at convergence, R2D2$_{\mathcal{A}_2,\mathcal{T}_2}$ achieves approximately $8$ times lower standard deviation for MRU. 
This trend underscores the superior robustness of R2D2$_{\mathcal{A}_2,\mathcal{T}_2}$ in handling variations in visibility weighting.

Panel (b) of Fig.~\ref{fig:uncertainty_figs} presents the reconstructed images obtained by R2D2$_{\mathcal{A}_1,\mathcal{T}_2}$ and R2D2$_{\mathcal{A}_2,\mathcal{T}_2}$, from the dirty images created with $\rho_\textrm{br}=1$ and $\rho_\textrm{br}=-1$ values for a selected RI simulation using the radio image 3C 353. The figure includes the ground truth, the dirty image for $\rho_\textrm{br}=1$, and reconstructed images of R2D2$_{\mathcal{A}_1,\mathcal{T}_2}$ and R2D2$_{\mathcal{A}_2,\mathcal{T}_2}$. 
It also presents their corresponding residual dirty images and relative uncertainty images $[\bsigma/\bmu](\widehat{\boldsymbol{X}})$.
Showcasing the results of the cases $\rho_\textrm{br}=-1$ and $\rho_\textrm{br}=1$ is motivated by the observed consistency in reconstruction quality across all Briggs parameter values.  These extremes represent uniform and natural weighting, effectively capturing the model’s robustness to visibility weighting variations. One can observe that the reconstructed images and mean image remain visually consistent across different $\rho_\textrm{br}$ values for both R2D2 models.  
The residual dirty images show discernible structures, particularly in the case of natural weighting ($\rho_\textrm{br} = 1$), which suggests that visibility weighting has a more noticeable impact on the fidelity to the dirty images than on the reconstructions themselves. Specifically, the residual dirty images of R2D2$_{\mathcal{A}_1,\mathcal{T}_2}$ exhibit ringing artifacts for $\rho_\textrm{br}=1$, which are absent in the corresponding residual dirty images of R2D2$_{\mathcal{A}_2,\mathcal{T}_2}$. 
The relative uncertainty images show comparable behavior for both R2D2 models across all $\rho_\textrm{br}$ variations, further confirming that the models deliver stable reconstructions despite changes in visibility weighting.

Fig.~\ref{fig:BR_study_BR_varation} provides a comprehensive quantitative analysis of the metric variations across different $\rho_\textrm{br}$ values. 
It depicts average values of SNR and logSNR values of R2D2 reconstructions as a function of Briggs parameter $\rho_\textrm{br}$. The standard deviation values are also provided at each $\rho_\textrm{br}$ point, shown as $\pm$~std annotations. R2D2$_{\mathcal{A}_1,\mathcal{T}_2}$ and R2D2$_{\mathcal{A}_2,\mathcal{T}_2}$ achieve their best image quality in terms of SNR at $\rho_\textrm{br}=0$, corroborating the fact that a balance between natural and uniform weighting often yields the highest reconstruction quality. In contrast, for logSNR, R2D2$_{\mathcal{A}_2,\mathcal{T}_2}$ achieves its peak value at $\rho_\textrm{br}=1$, whereas  R2D2$_{\mathcal{A}_1,\mathcal{T}_2}$ performs best at $\rho_\textrm{br}=0$. This suggests that R2D2$_{\mathcal{A}_2,\mathcal{T}_2}$ is more faithful to the standard expectation that natural weighting maintains optimal sensitivity and thus delivers higher dynamic range. The standard deviation values remain highly stable across variations of $\rho_\textrm{br}$, particularly for R2D2$_{\mathcal{A}_2,\mathcal{T}_2}$, and remain modest in comparison to the mean metrics.

While a similar study on robustness to visibility weighting could be conducted for benchmarking RI algorithms, the need for parameter fine-tuning and their relatively high computational cost make such a study impractical.

\begin{figure*}
    \centering
    \hspace{-1cm}
    \begin{tabular}{ll}
        \includegraphics[width=0.4\linewidth]{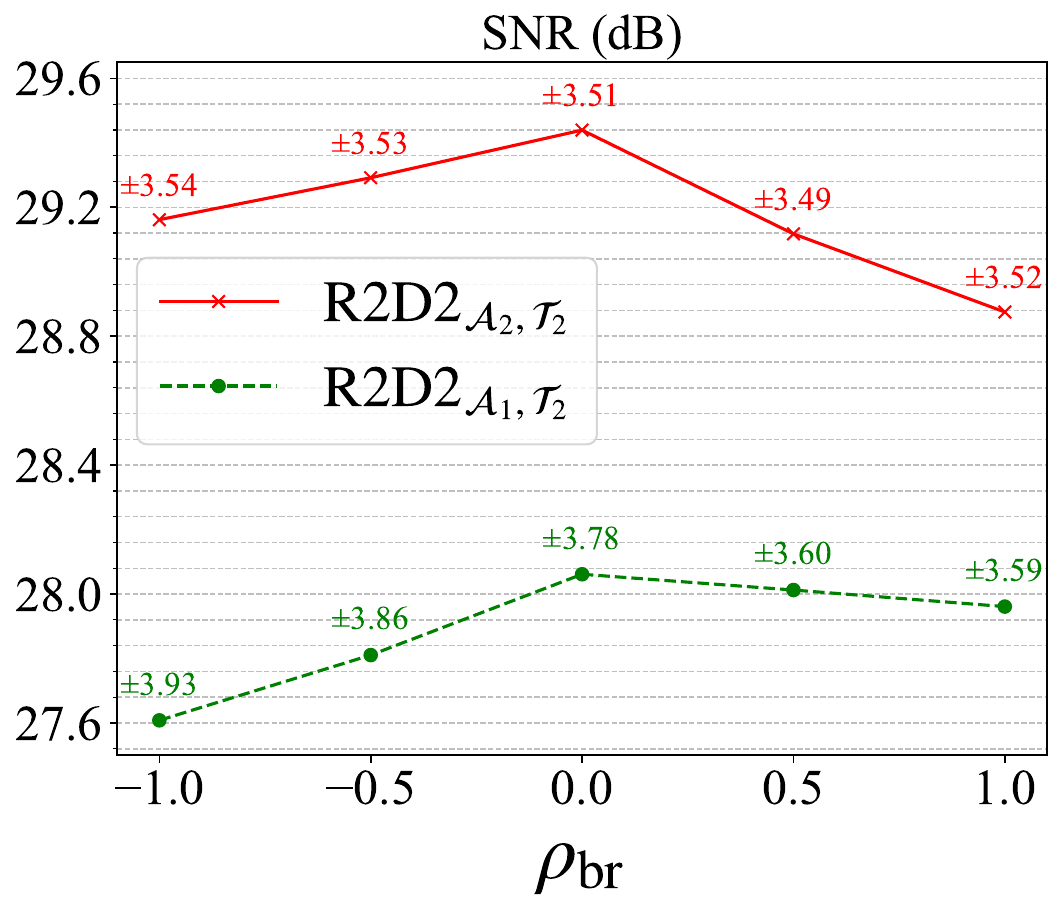} & 
        \includegraphics[width=0.4\linewidth]{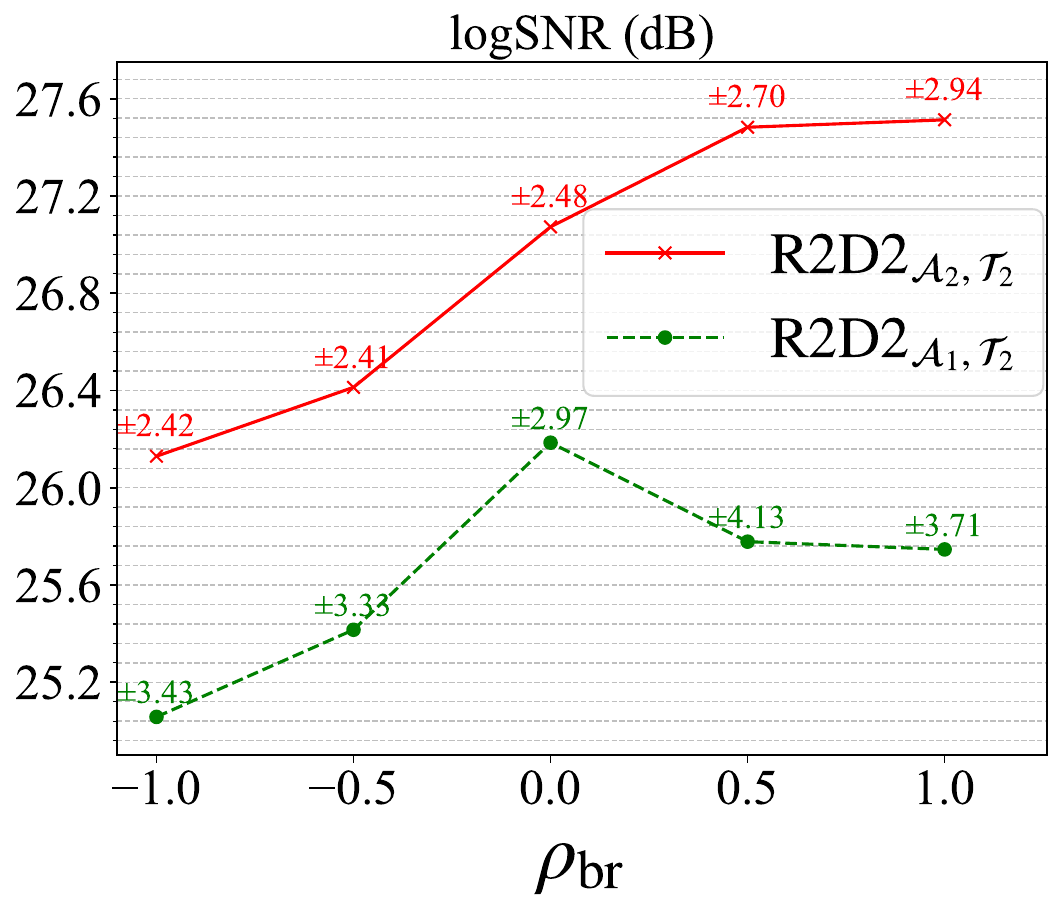} \\ 
        
    \end{tabular}
    \caption{Reconstruction results of R2D2 models in terms of SNR (left) and logSNR (right) shown as functions of Briggs parameter $\rho_\textrm{br}$. Each point represents the average metric calculated over 1000 inverse problems corresponding to a specific $\rho_\textrm{br}$ value. Standard deviation (±std) is reported at each point.}
    \label{fig:BR_study_BR_varation}
\end{figure*}


\section{Real data and results}\label{realdata}

In this section, we revisit VLA observations of the celebrated radio galaxy Cygnus~A. These data have been heavily scrutinized in recent works \citep[e.g.][]{dabbech2021, Arras21, Roth24} and most recently with the first incarnation of the R2D2 paradigm \citep{dabbech2024cleaning}, where full observational details can be found. We first highlight the impact of the generalized training set by comparing the performance of the new models R2D2$_{\mathcal{A}_1,\mathcal{T}_2}$ and R2D2$_{\mathcal{A}_2,\mathcal{T}_2}$ with the early model R2D2$_{\mathcal{A}_1,\mathcal{T}_1}$. We further analyse the performance of the new models to showcase the impact of the core DNN architecture U-WDSR on the image reconstruction quality. For a fair comparison with the early model R2D2$_{\mathcal{A}_1,\mathcal{T}_1}$, we adhered to the imaging settings of its training setup $\mathcal{T}_1$. We therefore formed images of size $N=512 \times 512$ with a pixel resolution corresponding to $\rho_\textrm{sr} = 1.5 $ using Briggs-weighted data with $\rho_\textrm{br}=0$. Under the convergence criterion defined in Section~\ref{series_convergence}, R2D2$_{\mathcal{A}_1,\mathcal{T}_1}$, R2D2$_{\mathcal{A}_1,\mathcal{T}_2}$, and R2D2$_{\mathcal{A}_2,\mathcal{T}_2}$ called for $12$, $16$, and $15$ iterations, respectively. 
To further investigate the proposed models' robustness, we examine epistemic uncertainty arising from  R2D2 realizations and variations of visibility weighting during imaging. To this aim, we generated $R=25$ reconstructions by combining all five model realizations (studied in Section \ref{uncertainty_realization}) and applying five different visibility weights during imaging through variations of the value of the Briggs parameter $\rho_{\textrm{br}}$ (studied in Section \ref{uncertainty_briggs}). Under this consideration, mean images are obtained by taking the pixel-wise mean of these $25$ reconstructions.

Reconstruction results are displayed in Fig.~\ref{fig:cynusA}. These include Cygnus~A reconstructions and associated residual dirty images obtained from selected realizations of R2D2$_{\mathcal{A}_1,\mathcal{T}_1}$, R2D2$_{\mathcal{A}_1,\mathcal{T}_2}$, and R2D2$_{\mathcal{A}_2,\mathcal{T}_2}$. 
Mean images $\bmu(\widehat{\boldsymbol{X}})$ and associated relative uncertainty images $[\bsigma/\bmu](\widehat{\boldsymbol{X}})$ obtained with R2D2$_{\mathcal{A}_1,\mathcal{T}_2}$ and R2D2$_{\mathcal{A}_2,\mathcal{T}_2}$ are also provided. For enhanced visual clarity, the residual dirty images are visualized on a linear scale, while all model estimate images and relative uncertainty images are displayed on a log$_{10}$ scale. 
Visual inspection suggests a general consistency of the reconstructions obtained with the different R2D2 models. Differences arise when examining faint emission with pixel values below $3$ orders of magnitude from the peak, such as the tails of the jets and the surrounding of the inner core of the radio galaxy (highlighted via a red ellipse).  
In particular, both R2D2$_{\mathcal{A}_1,\mathcal{T}_2}$ and R2D2$_{\mathcal{A}_2,\mathcal{T}_2}$ appear to succeed in capturing more fine-scale structure than R2D2$_{\mathcal{A}_1,\mathcal{T}_1}$. 
We first focus on U-Net models. Comparing individual realizations of R2D2$_{\mathcal{A}_1,\mathcal{T}_1}$ and R2D2$_{\mathcal{A}_1,\mathcal{T}_2}$ reveals that certain faint features are missing in one that are recovered in the other. More specifically, three features are highlighted using arrows. The blue arrows indicate recovered features, while red arrows mark those that were not recovered.  
R2D2$_{\mathcal{A}_1,\mathcal{T}_1}$ recovered two out of three features, the individual realization of R2D2$_{\mathcal{A}_1,\mathcal{T}_2}$ captured only one feature. This discrepancy is non-unexpected given the non-negligible uncertainty in these regions (see highlighted with black arrows in the relative uncertainty map). Interestingly, the mean image of R2D2$_{\mathcal{A}_1,\mathcal{T}_2}$ recovers one more feature than the individual realization ({i.e.}~two out of three). Their presence in the mean image is more reliable than in individual realizations, as it results from averaging over two sources of epistemic uncertainty, reducing the influence of model-specific variations.  
We then turn our attention to U-WDSR models. All three highlighted features are consistently recovered in the images corresponding to both  R2D2$_{\mathcal{A}_2,\mathcal{T}_2}$ realization and its mean image.  
The relative uncertainty images reveal a four-fold lower overall uncertainty for R2D2$_{\mathcal{A}_2,\mathcal{T}_2}$ compared to R2D2$_{\mathcal{A}_1,\mathcal{T}_2}$. This supports the observation that differences between R2D2$_{\mathcal{A}_2,\mathcal{T}_2}$ realizations and the mean image remain subtle. It also provides further confidence that all three features are real. Finally, analysis of the residual dirty images reveals a similar pattern of improvement across the three R2D2 models, where the new R2D2 models achieve higher data fidelity. Interestingly, the U-WDSR-based model enables a more homogenous residual structure than the U-Net-based R2D2 models, especially around the hotspots (highlighted in white circles), which are affected by calibration errors \citep{dabbech2021}. This observation is validated numerically by the lower values of the data fidelity metrics (reported inside the associated images). ~\\

\begin{figure*}
    \centering
    \setlength\tabcolsep{4pt}
    \renewcommand{\arraystretch}{0.0}
    \begin{tabular}{ccc}
    \begin{tikzpicture}[spy using outlines={circle, magnification=2, size=1cm}]
        \node[inner sep=0pt] (image) at (0,0) 
            {\includegraphics[width=0.39\linewidth]{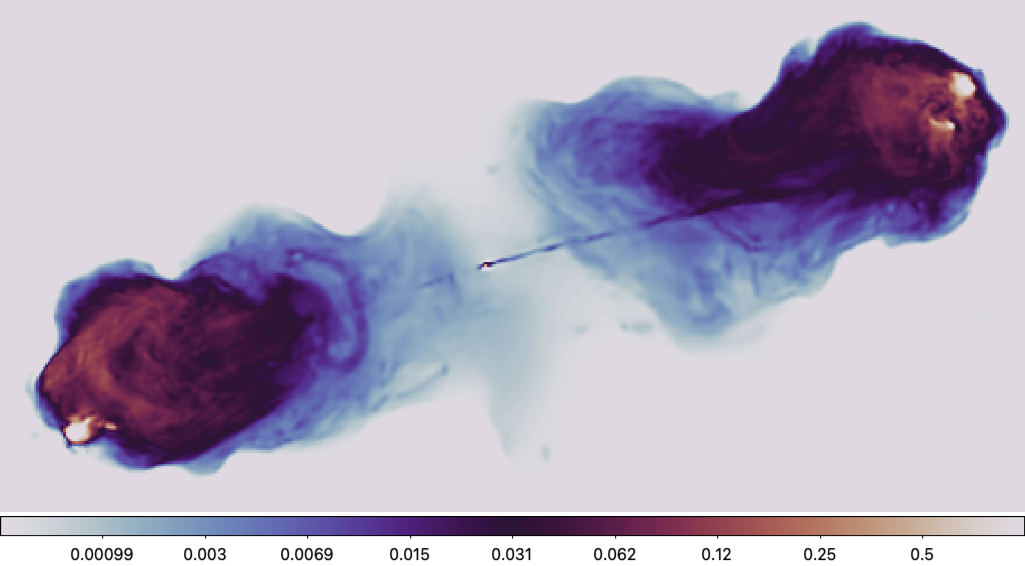}};
    
        \draw[dashed, thick, red] (image.center) ellipse (1.1cm and 1.4cm);

        \node at (-2.7,1.7) {\textcolor{black}{{$\widehat{\xb}_{\textrm{R2D2}_{\mathcal{A}_1,\mathcal{T}_1}}$}}};
    \end{tikzpicture} &
    \begin{tikzpicture}
    \begin{scope}
        \clip (2.3,-1.5) ellipse (1.3cm and 1.7cm);
        \node[scale=1.18] at (2.3,-1.5) 
            {\includegraphics[width=0.39\linewidth]{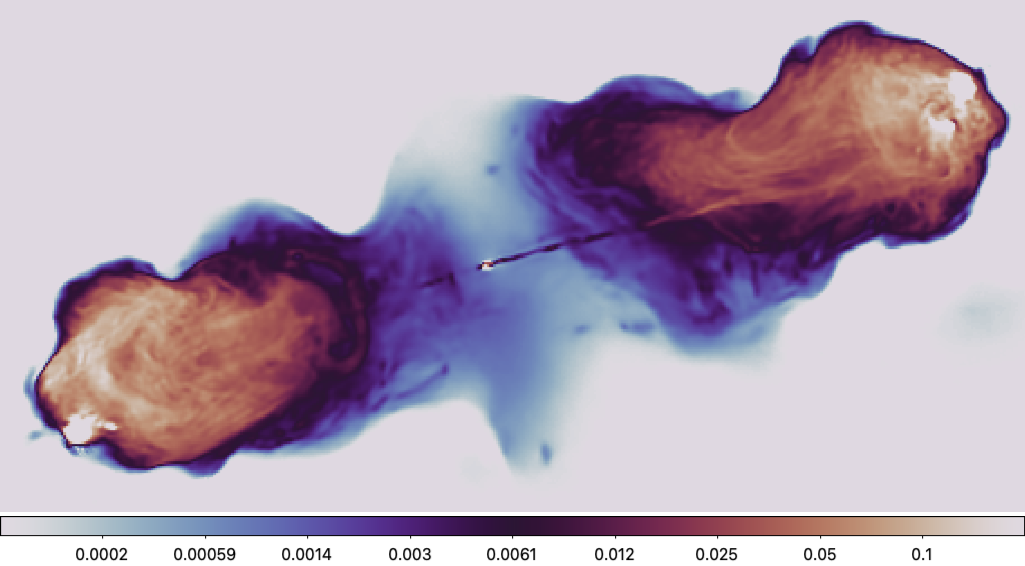}};
    \end{scope}
    \node at (2.3, -3.4) {\includegraphics[width=0.12\linewidth]{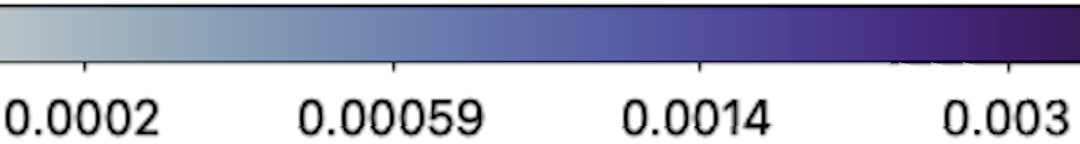}};
    \draw[->, blue, thick] (2.0975,0.02) -- (2.0975,-0.4);
    \draw[->, red, thick] (1.6, -0.23) -- (1.55, -0.55); 
    \draw[->, blue, thick] (3.1505, -2.445) -- (3.0425, -2.1075);
    \draw[red, thick, dashed] (2.3,-1.5) ellipse (1.3cm and 1.7cm); 
\end{tikzpicture} &
    \begin{tikzpicture}
        \node[inner sep=0pt] (image) at (0,0) {\includegraphics[width=0.39\linewidth]{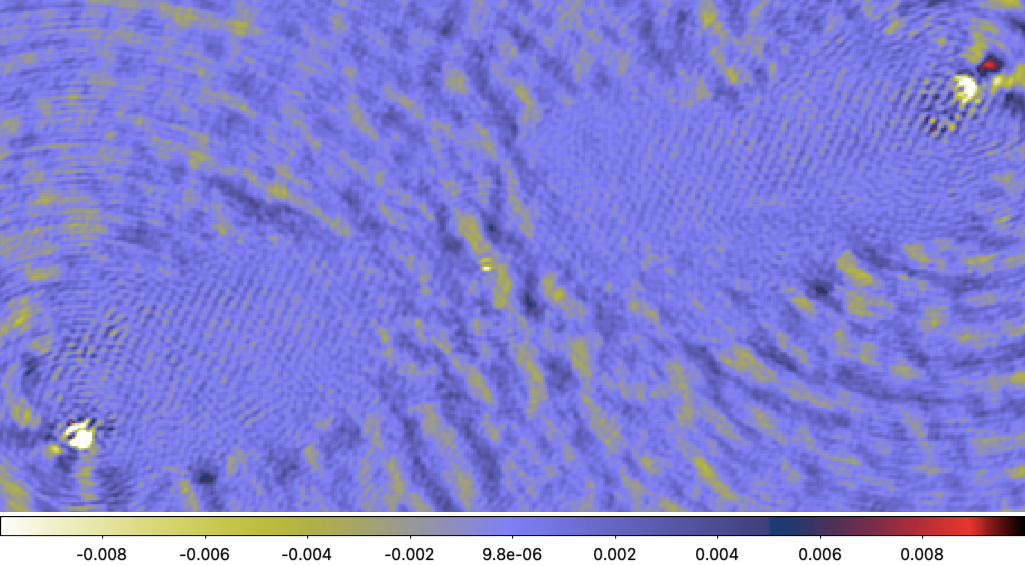}};
        \draw[dashed, thick, white] (-2.9,-1) ellipse (0.5cm and 0.5cm); 
        \draw[dashed, thick, white] (2.94,1.35) ellipse (0.5cm and 0.5cm); 
        \node at (-2.7,1.7) {\textcolor{white}{{$\widehat{\rb}_{\textrm{R2D2}_{\mathcal{A}_1,\mathcal{T}_1}}$}}};
        \node at (2.15,-1.35) {\textcolor{white}{$\textrm{RDR}$: 2.85 $\times 10^{-3}$}};
    \end{tikzpicture} \\
        \begin{tikzpicture}[spy using outlines={circle, magnification=2, size=1cm}]
            \node[inner sep=0pt] (image) at (0,0) {\includegraphics[width=0.39\linewidth]{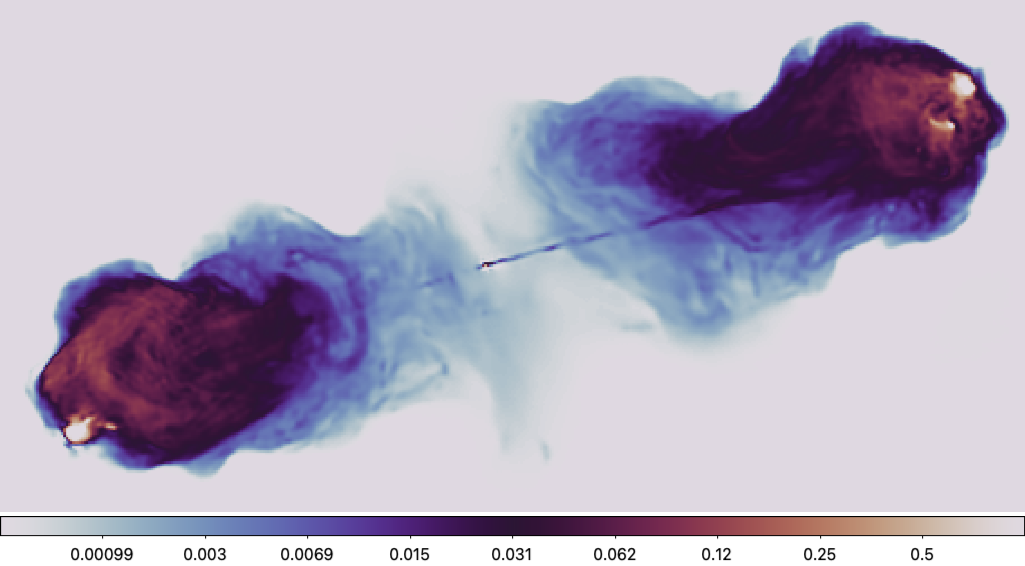}};
            \draw[dashed, thick, red] (image.center) ellipse (1.1cm and 1.4cm); 
            \node at (-2.7,1.7) {\textcolor{black}{{$\widehat{\xb}_{\textrm{R2D2}_{\mathcal{A}_1,\mathcal{T}_2}}$}}};
        \end{tikzpicture} &
        \begin{tikzpicture}
            \begin{scope}
                \clip (2.3,-1.5) ellipse (1.3cm and 1.7cm); 
                \node[scale=1.18] at (2.3,-1.5) 
                    {\includegraphics[width=0.39\linewidth]{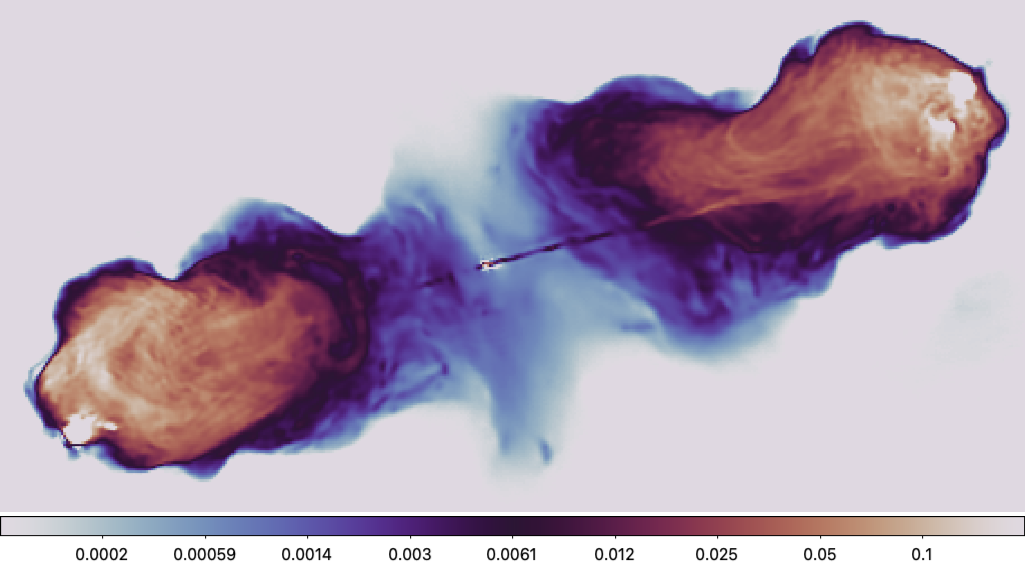}};
            \end{scope}
            \node at (2.3, -3.4) {\includegraphics[width=0.12\linewidth]{figs/real_data/zoom_colourbar.png}};
            \draw[red, thick, dashed] (2.3,-1.5) ellipse (1.3cm and 1.7cm); 
            \draw[->, red, thick] (2.0975,0.02) -- (2.0975,-0.4);
            \draw[->, blue, thick] (1.6, -0.23) -- (1.55, -0.55); 
            \draw[->, red, thick] (3.1505, -2.445) -- (3.0425, -2.1075);
        \end{tikzpicture} &
        \begin{tikzpicture}
            \node[inner sep=0pt] (image) at (0,0) {\includegraphics[width=0.39\linewidth]{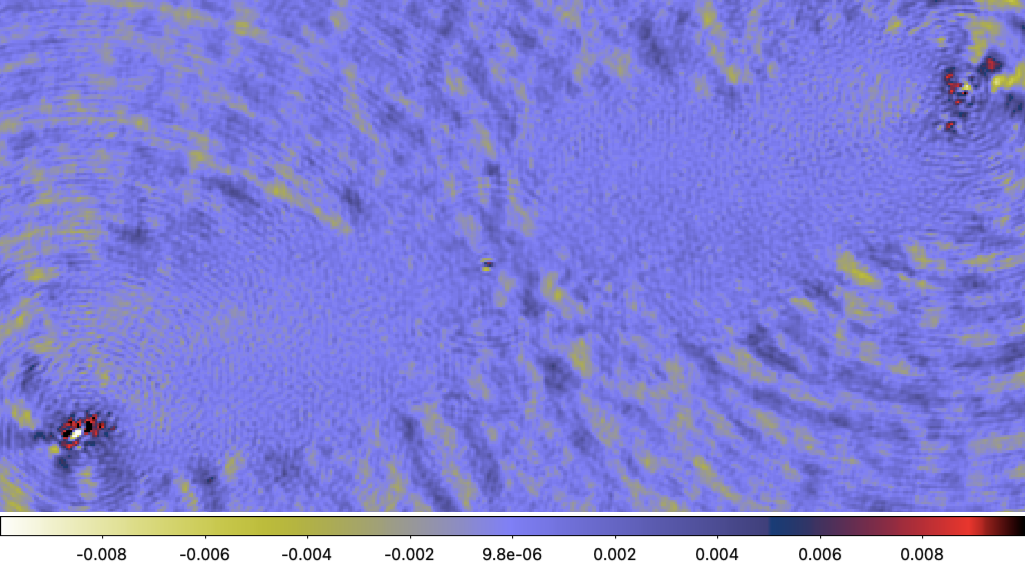}};
            \draw[dashed, thick, white] (-2.9,-1) ellipse (0.5cm and 0.5cm); 
            \draw[dashed, thick, white] (2.94,1.35) ellipse (0.5cm and 0.5cm); 
            
            \node at (-2.7,1.7) {\textcolor{white}{{$\widehat{\rb}_{\textrm{R2D2}_{\mathcal{A}_1,\mathcal{T}_2}}$}}};
            \node at (2.15,-1.35) {\textcolor{white}{{$\textrm{RDR}$: 2.68 $\times 10^{-3}$}}};
        \end{tikzpicture} \\
        
        \begin{tikzpicture}[spy using outlines={circle, magnification=2, size=1cm}]
            \node[inner sep=0pt] (image) at (0,0) {\includegraphics[width=0.39\linewidth]{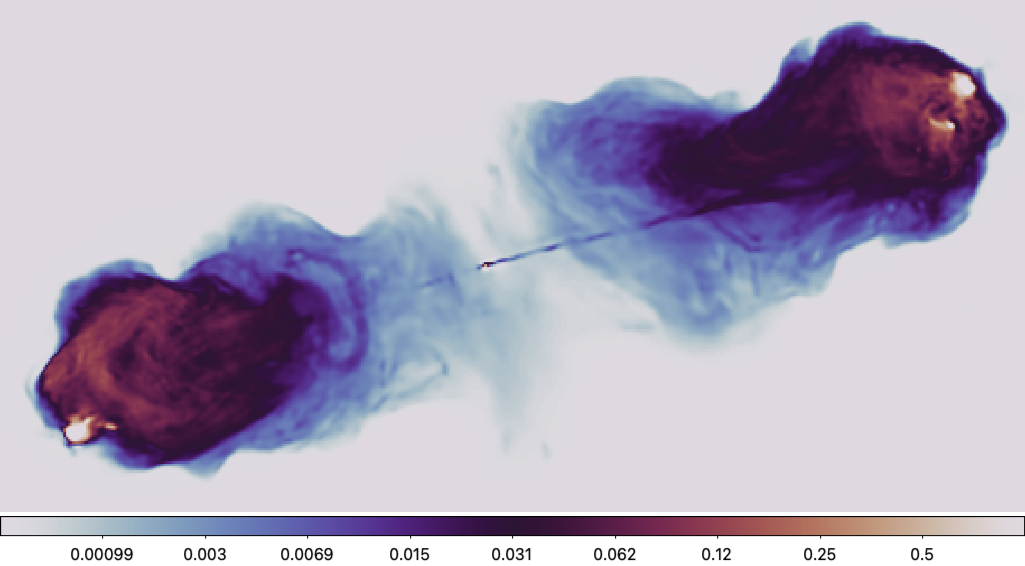}};
            \draw[dashed, thick, red] (image.center) ellipse (1.1cm and 1.4cm); 
            \node at (-2.4,1.6) {\textcolor{black}{{$\bmu(\widehat{\boldsymbol{X}}_{\textrm{R2D2}_{\mathcal{A}_1,\mathcal{T}_2}})$}}};
        \end{tikzpicture} &
        \begin{tikzpicture}
            \begin{scope}
                \clip (2.3,-1.5) ellipse (1.3cm and 1.7cm); 
                \node[scale=1.18] at (2.3,-1.5) 
                    {\includegraphics[width=0.39\linewidth]{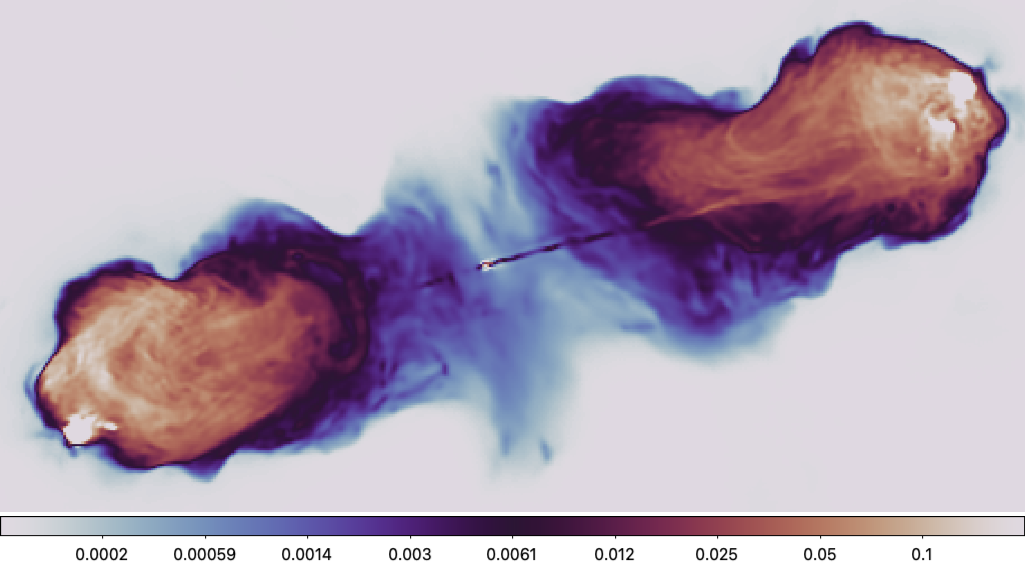}};
            \end{scope}
            \node at (2.3, -3.4) {\includegraphics[width=0.12\linewidth]{figs/real_data/zoom_colourbar.png}};
            \draw[red, thick, dashed] (2.3,-1.5) ellipse (1.3cm and 1.7cm); 
            \draw[->, blue, thick] (2.0975,0.02) -- (2.0975,-0.4);
            \draw[->, blue, thick] (1.6, -0.23) -- (1.55, -0.55); 
            \draw[->, red, thick] (3.1505, -2.445) -- (3.0425, -2.1075);
        \end{tikzpicture} &
        \begin{tikzpicture}
            \node[inner sep=0pt] (image) at (0,0) {\includegraphics[width=0.39\linewidth]{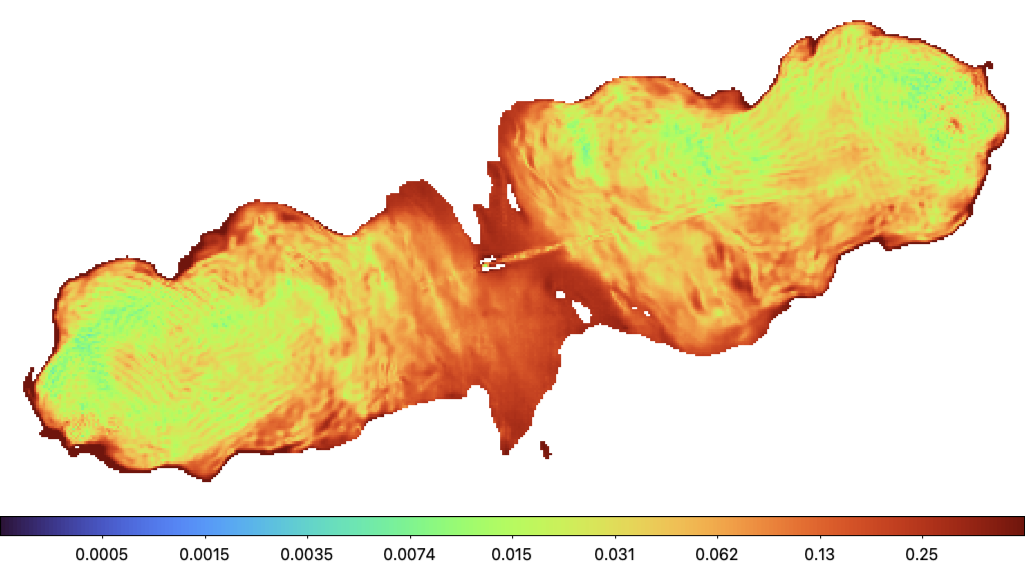}};
            \draw[dashed, thick, red] (image.center) ellipse (1.1cm and 1.4cm); 
            \node at (-2,1.6) {\textcolor{black}{{$[\bsigma/\bmu](\widehat{\boldsymbol{X}}_{\textrm{R2D2}_{\mathcal{A}_1,\mathcal{T}_2}})$}}};
            \node at (2.15,-1.35) {\textcolor{black}{{MRU$:8.98 \times 10^{-2}$}}};
            \draw[->, black, thick] (-0.15,1.2) -- (-0.15,0.95);
            \draw[->, black, thick] (-0.5,0.95) -- (-0.65,0.75);
            \draw[->, black, thick] (0.63,-0.7) -- (0.55,-0.45);
        \end{tikzpicture} \\
       
        \begin{tikzpicture}[spy using outlines={circle, magnification=2, size=1cm}]
            \node[inner sep=0pt] (image) at (0,0) {\includegraphics[width=0.39\linewidth]{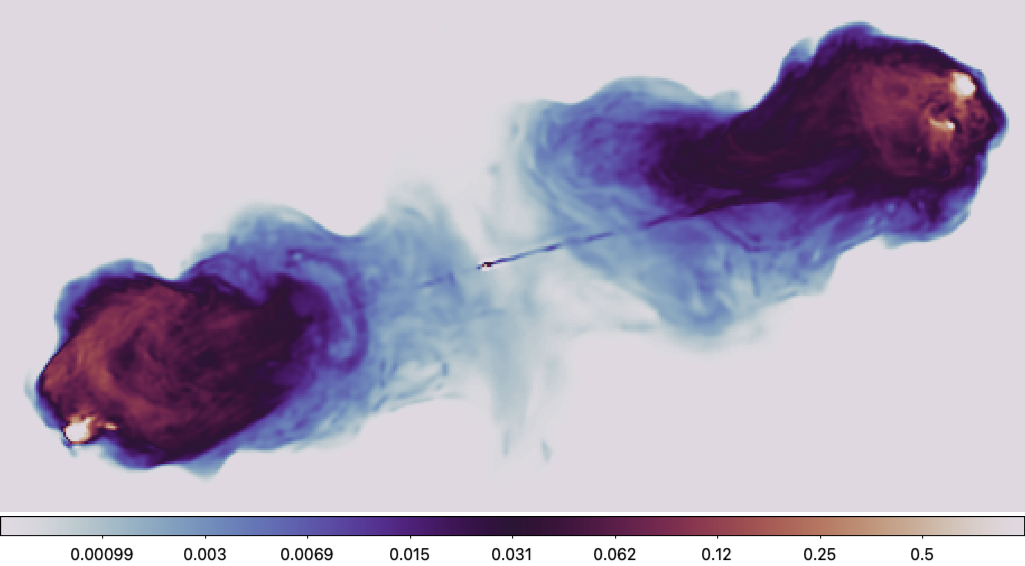}};
            \draw[dashed, thick, red] (image.center) ellipse (1.1cm and 1.4cm); 
            \node at (-2.7,1.7) {\textcolor{black}{{$\widehat{\xb}_{\textrm{R2D2}_{\mathcal{A}_2,\mathcal{T}_2}}$}}};
        \end{tikzpicture} &
        \begin{tikzpicture}
            \begin{scope}
                \clip (2.3,-1.5) ellipse (1.3cm and 1.7cm); 
                \node[scale=1.18] at (2.3,-1.5) 
                    {\includegraphics[width=0.39\linewidth]{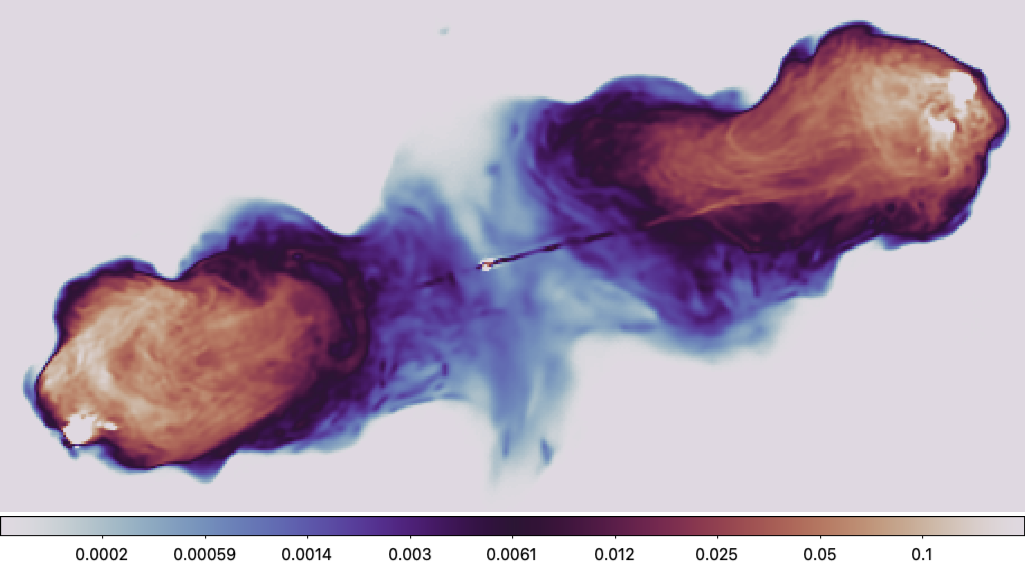}};
            \end{scope}
            \node at (2.3, -3.4) {\includegraphics[width=0.12\linewidth]{figs/real_data/zoom_colourbar.png}};
            \draw[red, thick, dashed] (2.3,-1.5) ellipse (1.3cm and 1.7cm); 
            \draw[->, blue, thick] (2.0975,0.02) -- (2.0975,-0.4);
            \draw[->, blue, thick] (1.6, -0.23) -- (1.55, -0.55); 
            \draw[->, blue, thick] (3.1505, -2.445) -- (3.0425, -2.1075);
        \end{tikzpicture} &
        \begin{tikzpicture}
            \node[inner sep=0pt] (image) at (0,0) {\includegraphics[width=0.39\linewidth]{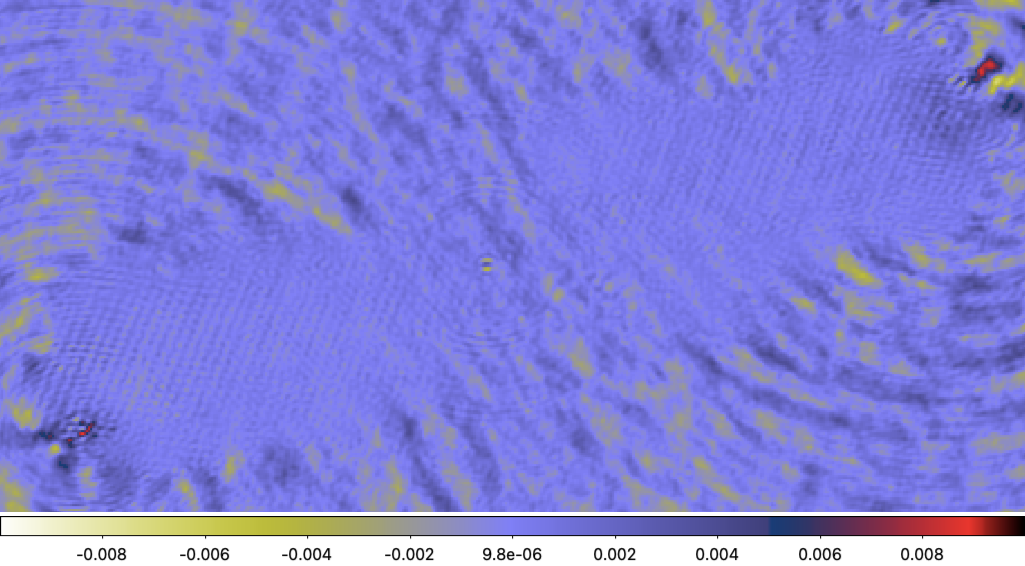}};
            \draw[dashed, thick, white] (-2.9,-1) ellipse (0.5cm and 0.5cm); 
            \draw[dashed, thick, white] (2.94,1.35) ellipse (0.5cm and 0.5cm);  
            \node at (-2.7,1.7) {\textcolor{white}{{$\widehat{\rb}_{\textrm{R2D2}_{\mathcal{A}_2,\mathcal{T}_2}}$}}};
            \node at (2.15,-1.35) {\textcolor{white}{{$\textrm{RDR}$: 2.46 $\times 10^{-3}$}}};
        \end{tikzpicture} \\
        
        \begin{tikzpicture}[spy using outlines={circle, magnification=2, size=1cm}]
            \node[inner sep=0pt] (image) at (0,0) {\includegraphics[width=0.39\linewidth]{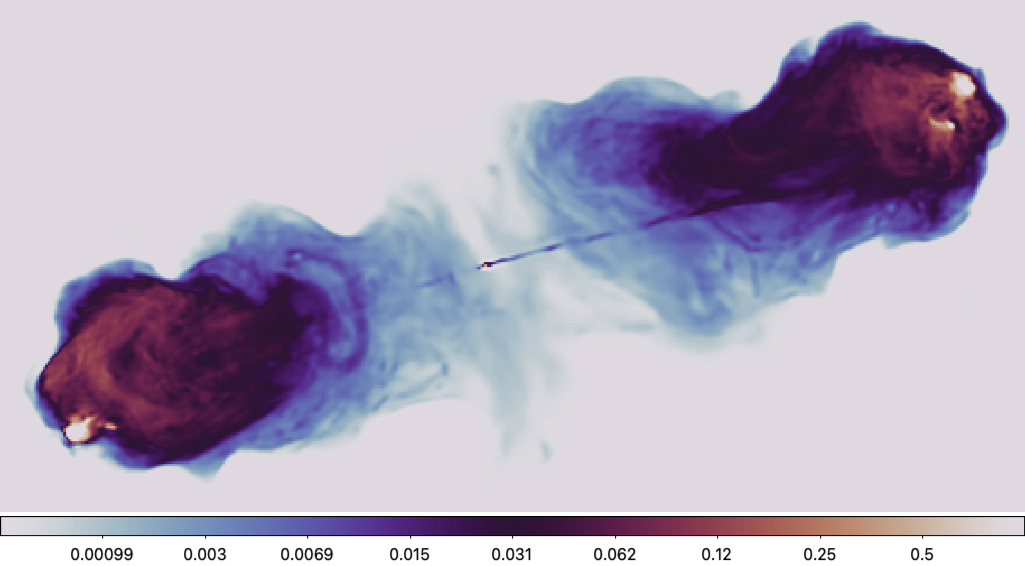}};
            \draw[dashed, thick, red] (image.center) ellipse (1.1cm and 1.4cm); 
            \node at (-2.4,1.6) {\textcolor{black}{{$\bmu(\widehat{\boldsymbol{X}}_{\textrm{R2D2}_{\mathcal{A}_2,\mathcal{T}_2}})$}}};
            
        \end{tikzpicture} &
        \begin{tikzpicture}
            \begin{scope}
                \clip (2.3,-1.5) ellipse (1.3cm and 1.7cm); 
                \node[scale=1.18] at (2.3,-1.5) 
                    {\includegraphics[width=0.39\linewidth]{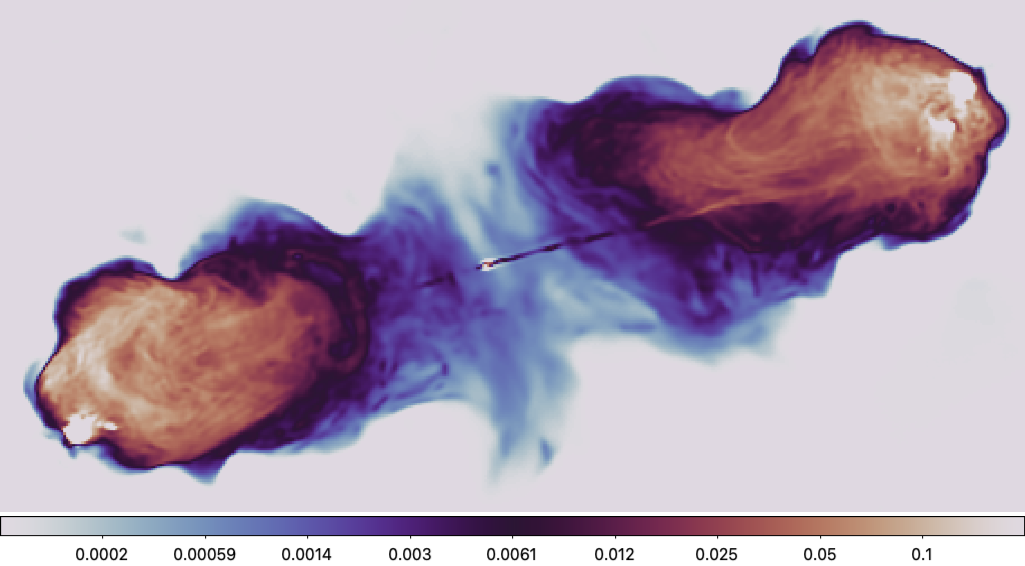}};
            \end{scope}
            \node at (2.3, -3.4) {\includegraphics[width=0.12\linewidth]{figs/real_data/zoom_colourbar.png}};
           \draw[->, blue, thick] (2.0975,0.02) -- (2.0975,-0.4);
            \draw[->, blue, thick] (1.6, -0.23) -- (1.55, -0.55); 
            \draw[->, blue, thick] (3.1505, -2.445) -- (3.0425, -2.1075);
            \draw[red, thick, dashed] (2.3,-1.5) ellipse (1.3cm and 1.7cm); 
        \end{tikzpicture} &
        \begin{tikzpicture}
            \node[inner sep=0pt] (image) at (0,0) {\includegraphics[width=0.39\linewidth]{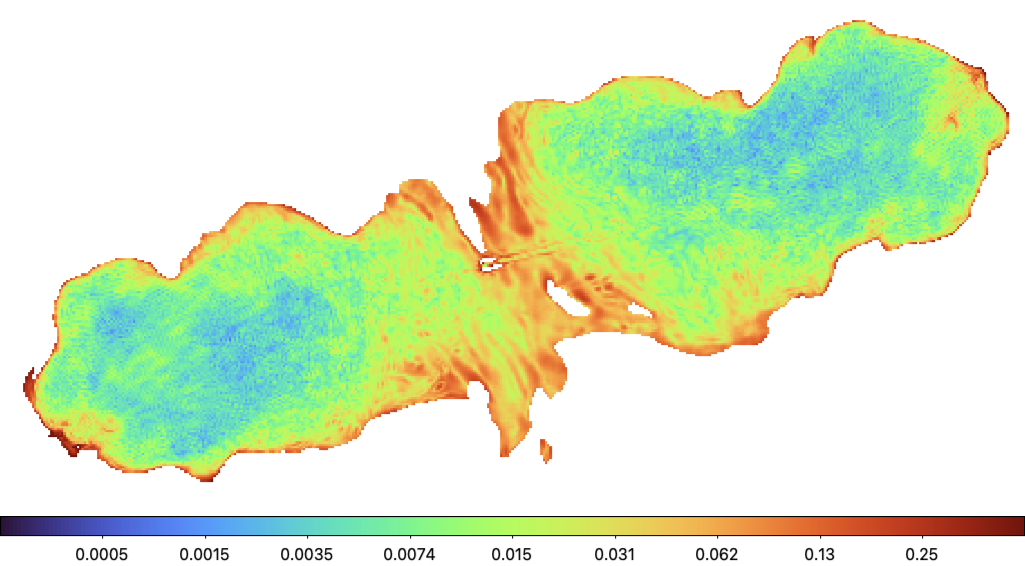}};
            \draw[dashed, thick, red] (image.center) ellipse (1.1cm and 1.4cm); 
            \node at (-2,1.6) {\textcolor{black}{{$[\bsigma/\bmu](\widehat{\boldsymbol{X}}_{\textrm{R2D2}_{\mathcal{A}_2,\mathcal{T}_2}})$}}};
            \node at (2.15,-1.35) {\textcolor{black}{{MRU$:2.37 \times 10^{-2}$}}};
            \draw[->, black, thick] (-0.15,1.2) -- (-0.15,0.95);
            \draw[->, black, thick] (-0.5,0.95) -- (-0.65,0.75);
            \draw[->, black, thick] (0.63,-0.7) -- (0.55,-0.45);
        \end{tikzpicture} \\
    \end{tabular}
    
    \caption{Cygnus~A reconstruction results using R2D2 models. The first (resp. second and fourth) row shows the image estimate (left) and corresponding residual dirty image (right) obtained with a given realization of R2D2$_{\mathcal{A}_1,\mathcal{T}_1}$ (resp. R2D2$_{\mathcal{A}_1,\mathcal{T}_2}$ and R2D2$_{\mathcal{A}_2,\mathcal{T}_2}$). The third (resp. fifth) row displays the mean image and associated relative uncertainty image delivered by R2D2$_{\mathcal{A}_1,\mathcal{T}_2}$ (resp. R2D2$_{\mathcal{A}_2,\mathcal{T}_2}$), and computed over $R=25$ reconstructions (five different model realizations combined with five different values of $\rho_\textrm{br}$). The red ellipse in the image estimates and the relative uncertainty image highlights the region of faint emission. The middle column visualizes a zoomed-in and contrast-enhanced view of this region. Coloured arrows highlight selected features,  in blue to point towards a recovered feature, and in red to indicate its location when missing (in black inside the relative uncertainty image). White circles in the residual dirty images indicate the locations of the hotspots. MRU and RDR metrics are reported inside the associated images.}
    \label{fig:cynusA}
\end{figure*}

\section{Conclusions} \label{Conclusions}

This paper revisits and significantly enhances the R2D2 algorithm robustness for RI imaging, specifically under the VLA observational setting targeting the formation of $512 \times 512$ monochromatic intensity images. These advancements span three key areas: training methodologies, convergence criteria, and DNN architecture. The generalized training set introduces stochastic variations, including randomization of the pixel resolution, visibility weighting parameter, sampling time, multinoise, and multiscan configurations—substantially improving the algorithm’s adaptability and robustness across diverse observational scenarios. 
To further enhance efficiency, a convergence criterion is introduced, whereby the reconstruction process is deemed
complete and iterations stop once the data residuals align with the noise level, rather than continuing until a fixed maximum number of DNNs is reached. This approach reduces computational cost during reconstruction. It also improves training efficiency by pruning converged inverse problems, allowing subsequent DNNs to focus on unsolved inverse problems,  leading to a more targeted optimization.
The core DNN architecture of the R2D2 algorithm is replaced with U-WDSR, a novel design, which offers enhanced imaging precision and improved robustness. 
The performance of the enhanced model R2D2$_{\mathcal{A}_2,\mathcal{T}_2}$ 
was rigorously validated through comprehensive simulation setups.
The results confirm that R2D2$_{\mathcal{A}_2,\mathcal{T}_2}$ consistently outperforms AIRI and uSARA in image reconstruction quality while achieving comparable data fidelity with significantly fewer iterations, resulting in a much faster reconstruction process. Furthermore, R2D2$_{\mathcal{A}_2,\mathcal{T}_2}$ exhibits much lower epistemic uncertainty compared to its U-Net-based counterpart, reaffirming the benefits of the U-WDSR architecture and the generalization strategies introduced in this work. This enhanced robustness holds across both sources of epistemic uncertainty, namely multiple R2D2 series realizations and variation in visibility-weighting schemes.
Illustration on real data consisting in VLA observations of Cygnus~A further validates the model’s effectiveness and robustness in accounting for the epistemic uncertainty, with the U-WDSR-based R2D2 model recovering finer details and achieving superior data fidelity compared to the U-Net-based models. 
While we did not include dedicated ablation studies isolating each individual factor, the effectiveness of the generalized training strategy is demonstrated through consistent performance gains across varied simulation setups and real data.

This work also introduces fully Python-based implementations of AIRI and uSARA, transitioning from MATLAB to a GPU-enabled Python framework. This transition not only improved computational efficiency but also enhanced the accessibility of the BASPLib library. 

Future work will focus on extending these advancements to address the challenges posed by large-scale imaging applications and adapting R2D2 for broader use cases. Specifically, future efforts will investigate (i) developing R2D2 for other telescopes or even a telescope-agnostic implementation, (ii) designing faceting strategies to enable seamless adaptation to any image size, including significantly larger dimensions, and (iii) generalizing the approach to wideband polarization imaging (iv) extending R2D2 to perform joint calibration and imaging. These developments will position R2D2 as a robust and scalable solution, paving the way for its integration into next-generation radio telescopes like the SKA and beyond.

\begin{acknowledgments} 
The authors would like to thank R.A. Perley for providing VLA observations of Cygnus A, and Adrian Jackson for his support related to Cirrus.
This research was supported by UK Research and Innovation through the EPSRC grant EP/T028270/1 and the STFC grant ST/W000970/1. The research was conducted using Cirrus, a UK National Tier-2 HPC Service at EPCC funded by the University of Edinburgh and EPSRC (EP/P020267/1). The National Radio Astronomy Observatory is a facility of the National Science Foundation operated under cooperative agreement by Associated Universities, Inc.
\end{acknowledgments}

\software{WSClean \citep{offringa2017},
 Meqtrees \citep{Noordam2010}, \href{http://localhost:63342/pythonProject/index.html}{BASPLib}, PyTorch \citep{paszke2019pytorch}, TorchKbNufft \citep{muckley:20:tah}, FINUFFT \citep{shih2021cuFINUFFT}, PyNUFFT \citep{lin2018python};
 }
\facility{VLA}

\section*{Data Availability}
R2D2 codes are available alongside AIRI and uSARA codes in the BASPLib code library on GitHub. BASPLib is developed and maintained by the Biomedical and Astronomical Signal Processing Laboratory (\href{https://basp.site.hw.ac.uk/}{BASP}). R2D2 DNN Series are available  in the data set at doi: \texttt{\href{https://doi.org/10.17861/e3060b95-4fe6-4b61-9f72-d77653c305bb}{10.17861/e3060b95-4fe6-4b61-9f72-d77653c305bb}}. 

Images used to generate training, validation, and testing datasets are sourced as follows. Optical astronomy images are gathered from NOIRLab/NSF/AURA/H.Schweiker/WIYN/T.A.Rector (University of Alaska Anchorage). Medical images are obtained from the NYU fastMRI Initiative database \citep{zbontar2018fastmri,knoll2020fastmri}. Radio astronomy images are obtained from the NRAO Archives, LOFAR HBA Virgo cluster survey~\citep{edler2023}, and LoTSS-DR2 survey~\citep{shimwell2022}. Observations of Cygnus A were provided by the National Radio Astronomy Observatory (NRAO; Program code: 14B-336). The self-calibrated data can be shared upon request to R.A. Perley (NRAO).

\bibliography{R2D2}{}
\bibliographystyle{aasjournal}

%
 



\end{document}